\documentclass{fundam}

\publyear{22}
\papernumber{2102}
\volume{185}
\issue{1}

\usepackage{url} 
\usepackage[ruled,lined]{algorithm2e}
\usepackage{graphicx}
\usepackage[lofdepth,lotdepth]{subfig}
\usepackage{caption} 
\usepackage{listings}
\usepackage{tikz}
\usetikzlibrary{automata, positioning, arrows}

\begin{document}

\title{Towards a Self-Replicating Turing Machine}

\address{ralph.lano@th-nuernberg.de}

\author{Ralph P. Lano\thanks{I would like to thank W.Z. Taylor for discussions, feedback and suggestions.}\\
Technische Hochschule Nürnberg - Georg Simon Ohm\\ Keßlerplatz 12, 90489 Nürnberg, Germany\\
ralph.lano@th-nuernberg.de}

\maketitle

\runninghead{Ralph P. Lano}{Towards a Self-Replicating Turing Machine}

\begin{abstract}
We provide partial implementations of von Neumann's universal constructor and universal copier, starting out with three types of simple building blocks using minimal assumptions.  Using the same principles, we also construct Turing machines.  Combining both, we arrive at a proposal for a self-replicating Turing machine.  Our construction allows for mutations if desired, and we give a simple description language.
\end{abstract}
\begin{keywords}
self-replication, Turing machine, universal constructor, von Neumann, self-reproduction, artificial life, nanorobots
\end{keywords}

\section{Introduction}

Long before the details of biological self-reproduction were understood, von Neumann proposed a purely logical model for self-reproduction \cite{neumann1966theory}.  It consisted of a universal constructor automaton A, a universal copier automaton B, and descriptions thereof, $\Phi(A)$ and $\Phi(B)$.  It also included a third automaton C that would control A and B, and might be needed for additional manipulations.  Finally, he introduced an automaton D that could be any automaton.  Automaton D had nothing to do with self-reproduction: however, it could undergo mutation \cite{rocha2015neumann}.  

Clearly, living organisms provide an implementation of von Neumann's scheme.  The copier automaton B can be associated with the bacterial replisome \cite{o2013principles}, which makes copies of DNA.  A somewhat simpler version is the protein RNA polymerase, which in the process called \textit{transcription}, synthesises RNA from DNA, basically making a negative copy.  The universal constructor automaton A can be associated with the ribosomal proteins, which take information from RNA and build proteins.  This process is called \textit{translation}.  And finally, what von Neumann called the description of an automaton is what biologists call the genetic code \cite{alberts2017molecular}.

Self-reproducing structures have been extensively presented and discussed with various levels of detail in the literature \cite{merkle2004kinematic}.  In this paper, we would like to give a "mechanical" toy implementation of von Neumann's automata.  Since our implementation draws inspiration from biological systems, we would like to review a few useful facts about biological systems.

The basic building blocks of biological systems are the amino acids.  There is a total of twenty amino acids that are used by living cells.  It is essential to realize that the amino acids serve a dual purpose: on the one hand, they encode genetic information, but on the other hand, they also form the basic building blocks for proteins.  Four of the amino acids, the nucleotides (A, C, G, and T), are special in the sense that in addition to being used for construction purposes, they are also used to encode the genetic code. 

The genetic code, that is RNA or DNA, is encoded in codons.  Every codon consist of three nucleotides (A, C, G, and T), and each  codon corresponds to one of the twenty amino acids.  In addition there are start and stop codons.  For every codon there is an exact match, the corresponding anticodon.  

In the process of transcription, RNA polymerase transcribes parts of the genetic code into short strands of messenger RNA (mRNA).  This mRNA is then translated by the ribosome into construction information.  For this it uses transfer RNA (tRNA).  Transfer RNA has an anticodon on one end and one of the twenty amino acids on the other end.   In the process of translation the anticodon of the tRNA is matched up with the codon of the mRNA, and the amino acid at the end of the tRNA is then added to form shorter and longer chains of amino acids, which are also called peptides or polypeptides.  The order of the codons in the mRNA dictates the order in which amino acids are added to the polypeptide chain.  Long polypeptide chains are also called proteins.  Proteins form one-dimensional, two-dimensional and three-dimensional structures.  Two very special proteins are RNA polymerase and ribosomes.

\section{Assumptions}

With this somewhat oversimplified picture of biological processes in the back of our mind, let us state our assumptions about our toy mechanism.  We want to be able to build arbitrary 3D structures out of little three-dimensional cubes, which we will call "blocks".  The building instructions should come from some "genetic code", which itself should be expressible with these blocks.  Also, the constructor, the copier and any other automaton should be made out of these blocks.
The blocks live on a three-dimensional grid, we assume there are no gravitational forces and, for simplicity, all have the same size.  They are our fundamental building units and there is basically an unlimited supply of them.  There are three different types of blocks and in the following we will show that they suffice to produce self-reproducing behavior.
\begin{itemize}
\item \textbf{Normal blocks:} Can be glued to other blocks, including mover and gluer, but usually are inert.  The gluing is initiated from the outside by the gluer.  Once glued they will stick together and form \textit{compounds}.  They can also be unglued from the outside for reuse.
\item \textbf{Mover blocks:} Can expand to twice their size and after a fixed amount of time, they will contract again.  They expand in only one direction.  For our simulations, the expansion is triggered by a global timer, however, it could also be triggered through several other means.  They can move a significant amount of other blocks, about ten to twenty.  For construction or during transport, they might need to be in an inactive state, but we do not worry about this.  However, a critical assumption is that movers can not break blocks apart that are glued together.
\item \textbf{Gluer blocks:} If another block gets next to the gluer, that block goes into the glue state.  If another block is close, they get glued together.  After some time a block returns to its normal state, but stays glued.  The gluing has direction.  The gluer can be in an inactive state, which may be needed during construction and transport.  The gluing may need energy or it maybe catalytic.
\end{itemize}

\begin{figure}[h]
\centering
\subfloat[][]{
\includegraphics[scale=0.5]{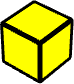}
\label{fig2_1:sf1}}
\qquad
\subfloat[][]{
\includegraphics[scale=0.5]{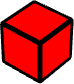}
\label{fig2_1:sf2}}
\qquad
\subfloat[][]{
\includegraphics[scale=0.5]{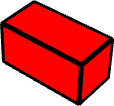}
\label{fig2_1:sf3}}
\qquad
\subfloat[][]{
\includegraphics[scale=0.5]{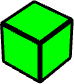}
\label{fig2_1:sf4}}
\caption{Different block types: (a) normal, 
(b) mover, (c) mover expanded, and (d) gluer.}
\label{fig2_1:fig}
\end{figure}

For our simulations we represent the different kinds of blocks through different colors (Figure \ref{fig2_1:fig}).  Normal blocks are yellow, mover blocks are red, and glue blocks are green.  In addition, we use mostly a blue color to indicate compounds.

Comparing our toy model with the genetic world, our blocks are similar to the amino acids.  Compounds can be considered to be equivalent to polypeptides or proteins.  The gluing corresponds to a strong chemical bond, hence the function of the gluer is to initiate that strong chemical bond.  The assumptions that go into the mover block are critical and non-trivial.  Clearly, it requires some form of energy source.  
We assume that in our world there is no gravity, either because there is none, or because our blocks are floating in some form of fluid.  Besides the strong bond, some weak, van der Waals type attractive force between blocks might be helpful.
Also, let us point out, that our blocks need not be biological, they could also be chemical, mechanical, or electromechanical, for instance.  The glue could actually be real glue, or it could be some mechanical connection.  The expansion of the mover could be caused biologically, chemically, electromagnetically, or hydraulically.  However, we prefer to think of the blocks as being as trivial as possible.

\section{Codons, Anticodons, mRNA and tRNA}

We start with the construction of our genetic code.  It should be expressible with our blocks.  In the genetic analogon, what we need are codons and anticodons.

\subsection{Codons and Anticodons}
Codons and their respective anticodons are kind of like key and lock: they need to match each other to fit.  There also should be a one-to-one correspondence between them.  We build codons and anticodons out of normal blocks that are glued together. Codons can be of different lengths.  We call them 2-codons because they are two blocks wide.

\subsubsection{2-Codon and 2-Anticodon}
The simplest useful codon is made out of three blocks.  On the bottom are the two possible 2-codons, on the top the matching 2-anticodons.
\begin{table}[h!]\small
  \begin{center}
  \begin{tabular}{cc}
\includegraphics[scale=0.25]{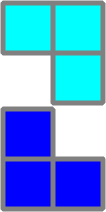} & \includegraphics[scale=0.25]{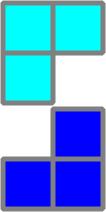}  \\
A &	B \\
  \end{tabular}
\end{center}
\end{table}

\subsubsection{3-Codon and 3-Anticodon}
The next simplest codon is the 3-codon.  On the bottom are the 3-codons, on the top the 3-anticodons.  Notice, that 3-codons and 3-anticodons have different shapes.
\begin{table}[h!]\small
  \begin{center}
  \begin{tabular}{ccc}
\includegraphics[scale=0.25]{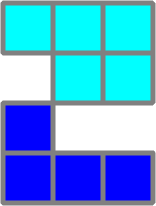} & \includegraphics[scale=0.25]{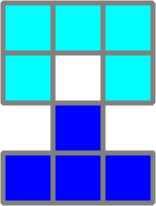} & \includegraphics[scale=0.25]{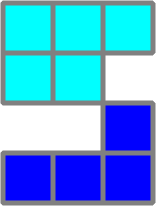}  \\
A &	B & C\\
  \end{tabular}
\end{center}
\end{table}
It is important to realise that there should be no mismatching possible.  Consider the example below.
\begin{table}[h!]\small
  \begin{center}
  \begin{tabular}{cc}
\includegraphics[scale=0.25]{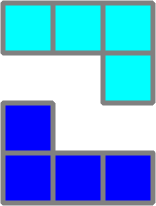}  \\
Mutation. \\
  \end{tabular}
\end{center}
\end{table}
If the 3-anticodons would have the same shape as the 3-codons, a possible mismatch could happen: codon A and codon B could match with the given anticodon.  In general, this is undesirable, unless we want to allow for mutations.

\subsubsection{4-Codon and 4-Anticodon}
As for the 4-codon, codons and anticodons are symmetrical again, and there is a total of six 3-codons, which is mathematically 2 out of 4 combinations.
\begin{table}[h!]\small
  \begin{center}
  \begin{tabular}{cccccc}
\includegraphics[scale=0.25]{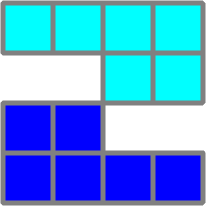} & \includegraphics[scale=0.25]{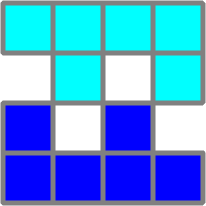} & \includegraphics[scale=0.25]{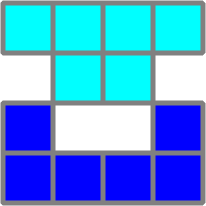} & \includegraphics[scale=0.25]{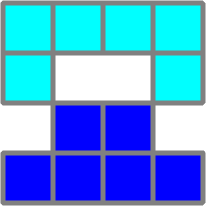} & \includegraphics[scale=0.25]{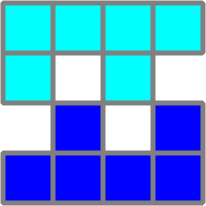} & \includegraphics[scale=0.25]{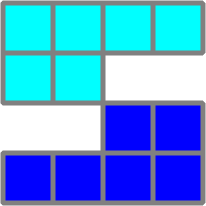}   \\
A &	B & C & D & E & F \\
  \end{tabular}
\end{center}
\end{table}
In general, there will be n/2 out of n combinations, hence the length of the codon is determined by the amount of information we want to store in the basic codon.  For our purposes, four will suffice.

\subsubsection{Mutations}
Usually, we want the matching to be exact with no room for mutations.  However, it would be easy to introduce mutations, if we were to allow anticodons that are not unique, like what we have seen in the 3-codon example above or the following mutating 4-anticodon.
\begin{table}[h!]\small
  \begin{center}
  \begin{tabular}{cc}
\includegraphics[scale=0.25]{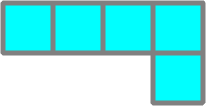}  \\
Mutation anticodon.\\
  \end{tabular}
\end{center}
\end{table}
In a later match, it could result in three possible codon matches, and hence has a two out of three chance to introduce a mutation.  By regulating the number of these mutating anticodons, we could modify the rate of mutation.  A brief note on wording: as  \cite{merkle2004kinematic} suggested, self-replication means replication without error, and self-reproduction means replication with error.

\subsection{RNA}
Placing several codons in a chain results in a strand of RNA.  If we want them to stay together, we place them inside some encasing.
\begin{figure}[h]
\centering
\includegraphics[scale=0.25]{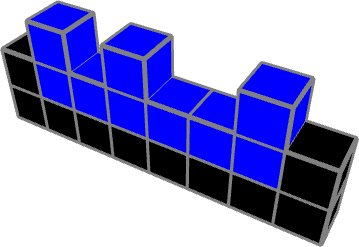}
\caption{RNA with encasing.}
\label{fig3_6:fig}
\end{figure}
Depending on the information content we need, we can choose our basic building blocks to be 2-codons, 3-codons, or n-codons as needed.  For everything we want to show, 4-codons suffice.  Notice, that the RNA has a direction: it could be read from left-to-right or from right-to-left.  We will not worry about this here, but introducing a start-codon solves the problem, or giving the codons an asymmetric shape is also a solution.

\subsection{Matching and Copying}
\begin{figure}[h]
\centering
\subfloat[]{
\includegraphics[scale=0.25]{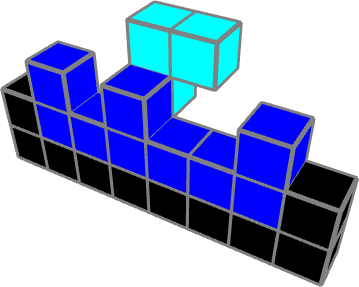}
\label{fig3_7:sf1}}
\qquad
\subfloat[]{
\includegraphics[scale=0.25]{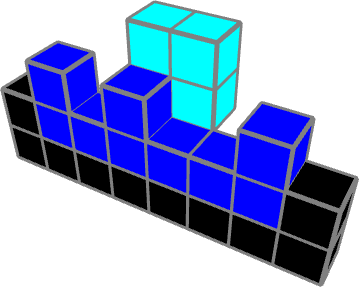}
\label{fig3_7:sf2}}
\caption{Matcher: (a) no match, (b) match.}
\label{fig3_7:fig}
\end{figure}

A central mechanism that we need again and again is the matcher.  The task of the matcher is to match codon with anticodon.  Consider Figure \ref{fig3_7:fig}, 
where we have in dark blue an RNA strand made out of three 2-codons.  We see an anticodon in light blue, originally positioned behind the RNA strand.  This could be an arbitrary anticodon, i.e., it does not necessarily match with the codon in the RNA.  We push the anticodon from behind.  If it matches, it will fit, if it does not match, it will not fit.  If it does not fit, we discard it, and pick the next one.  Once we have a match, we move the RNA by one codon length to the left.  Thus, we are able to uniquely match codons and anticodons this way.  This process will work for any of our n-codon/anticodon pairs.
Having a matcher, the next step is obvious: the copier.  It uses the matcher to create a "negative" of the RNA strand, made up entirely of anticodons.  If we repeat this step, we have a copy of the original.  In the genetic world this process is performed by the RNA polymerase and the process is called transcription when messenger RNA is created this way.

\subsection{Transfer RNA}
Next, we would like to consider the process of translation, that is protein synthesis.  To see how that might come about, we need to take a closer look at transfer RNA (tRNA).
\begin{table}[h!]\small
  \begin{center}
  \begin{tabular}{cc}
\includegraphics[scale=0.25]{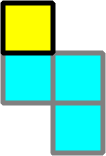} & \includegraphics[scale=0.25]{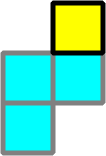}  \\
A &	B \\
  \end{tabular}
\end{center}
\end{table}
On the lower part, it looks like an anticodon (light blue) and on the top part we could put anything we want.  In the example shown, it is a normal block (yellow), on the left for anticodon A and on to the right for anticodon B.

\subsection{Building}
In the translation process, which we prefer to call \textit{building}, we first follow the matching steps outlined above.  Consider Figure \ref{fig3_9:fig}.  We have messenger RNA in an encasing consisting of the codons "AAB".  Initially, the tRNA are behind the mRNA.  For convenience, the tRNA is arranged in the correct order, but that is not essential.  In step 2, we push the tRNA forward, which is possible since the codon and anticodon parts match.
\begin{figure}[h]
\centering
\subfloat[Step 1]{
\includegraphics[scale=0.25]{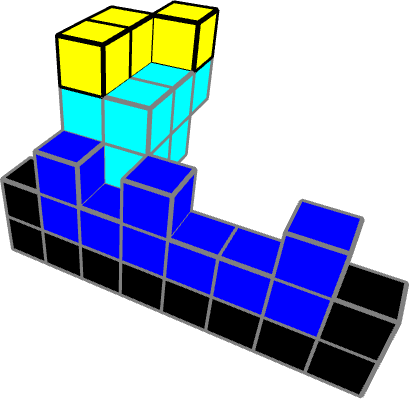}
\label{fig3_9:sf1}}
\qquad
\subfloat[Step 2]{
\includegraphics[scale=0.25]{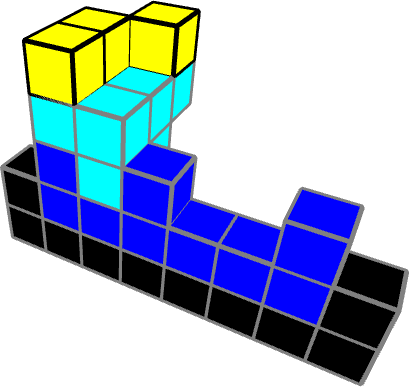}
\label{fig3_9:sf2}}

\subfloat[Step 3]{
\includegraphics[scale=0.25]{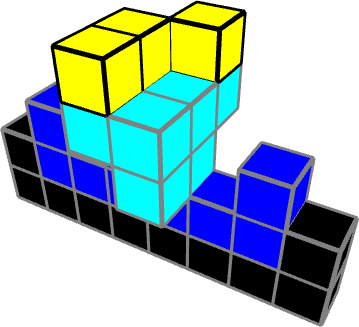}
\label{fig3_9:sf3}}
\qquad
\subfloat[Step 4]{
\includegraphics[scale=0.25]{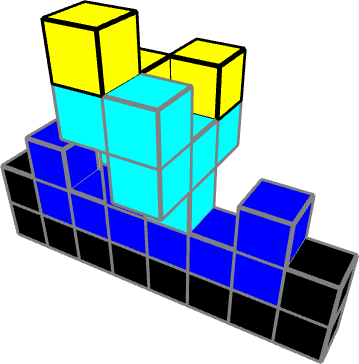}
\label{fig3_9:sf4}}
\end{figure}
\setcounter{figure}{3} 
\setcounter{subfigure}{4}
\begin{figure}[h]
\centering
\subfloat[Step 5]{
\includegraphics[scale=0.25]{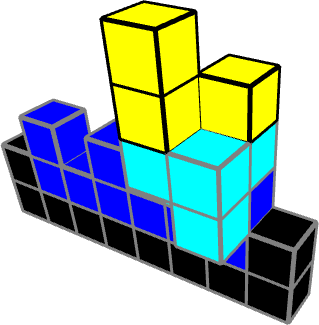}
\label{fig3_9:sf5}}
\qquad
\subfloat[Step 6]{
\includegraphics[scale=0.25]{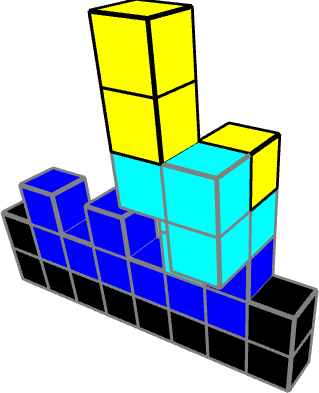}
\label{fig3_9:sf6}}
\subfloat[Step 7]{
\includegraphics[scale=0.25]{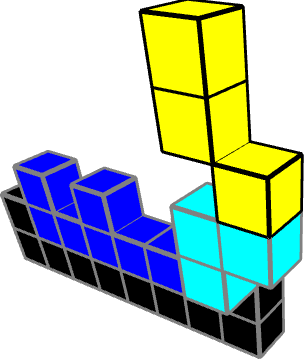}
\label{fig3_9:sf7}}
\qquad
\subfloat[Step 8]{
\includegraphics[scale=0.25]{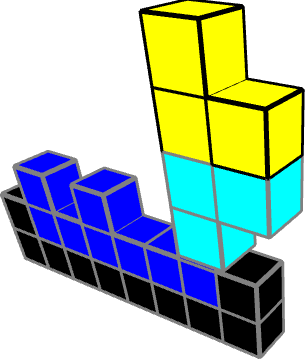}
\label{fig3_9:sf8}}
\caption{Simple builder}
\label{fig3_9:fig}
\end{figure}
In step 3, we move the mRNA by two to the left, and match the next codon anticodon pair.  Next in step 4, we push the first of the tRNA up by one.
Again, we move the mRNA by two to the left, and match the next codon anticodon pair in step 5, and push the second tRNA forward.  We also discard the anticodon part of the first tRNA, leaving the normal block that the first tRNA was carrying.  In step 6, we push the second tRNA up by one.  In step 7, we move the mRNA by two to the left again, but there are no more anticodons to match.  We push the third tRNA forward.  In the last step, we push the third tRNA up by one.
After discarding the anticodon parts, we are left with a new piece (yellow), which happens to resemble an "A 2-codon".  Hence the instruction "AAB" produces an "A 2-codon".  We call "AAB" the builder description language (BDL).
With this builder "machine" we can build basically any 2-by-n two-dimensional structure.  Below is a list of what we can build with three 2-codons, listing all possibilities.
\begin{table}[h!]\small
  \begin{center}
  \begin{tabular}{cccccccc}
\includegraphics[scale=0.25]{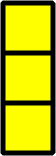} & \includegraphics[scale=0.25]{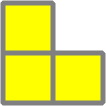} & \includegraphics[scale=0.25]{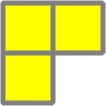} & \includegraphics[scale=0.25]{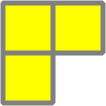} & \includegraphics[scale=0.25]{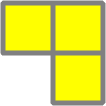} & \includegraphics[scale=0.25]{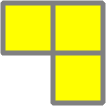} & \includegraphics[scale=0.25]{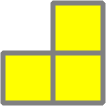} & \includegraphics[scale=0.25]{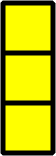}  \\
AAA  & AAB 	 & ABA 	 & BAA 	 & ABB 	 & BAB 	 & BBA 	 & BBB \\
  \end{tabular}
\end{center}
\end{table}

We have just shown how one could encode information in a type of genetic code made out of our basic building blocks, and we motivated how the process of matching, copying and building could be performed with these basic ingredients.

\subsection{3D Structures}
\begin{figure}[h]
\centering
\includegraphics[scale=0.2]{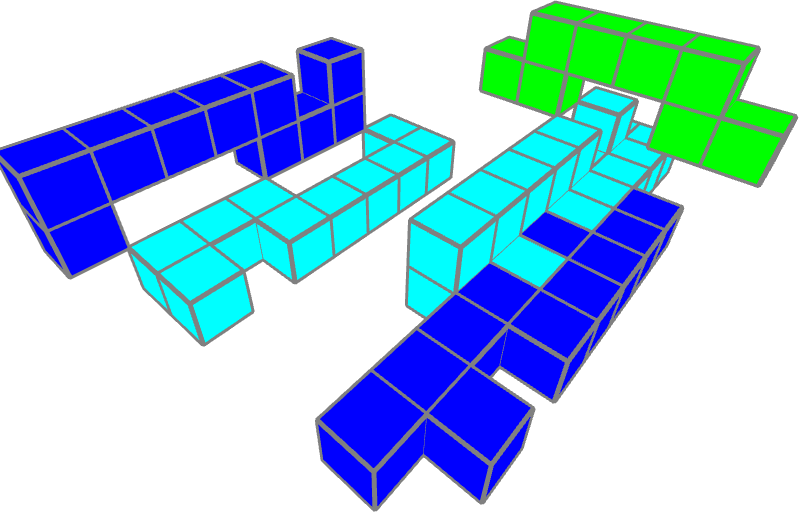}
\caption{Several interlocking 2-by-n structures.}
\label{fig5_27:fig}
\end{figure}

Being able to build only 2-by-n structures might seem useless, if one wants to build three-dimensional structures.  To convince ourselves that that is not so, consider the following two-dimensional structures, and how they can be made to interlock to build complex three-dimensional structures (Figure \ref{fig5_27:fig}).

\begin{figure}[h]
\centering
\subfloat[Before assembly]{
\includegraphics[scale=0.25]{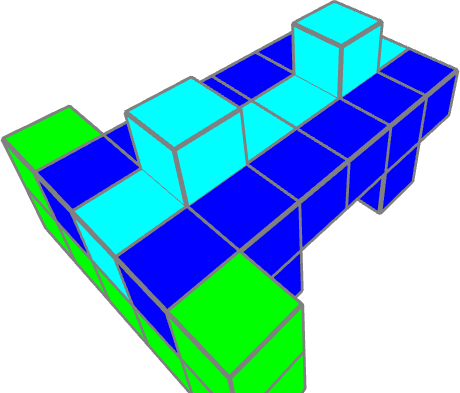}
\label{fig5_28:sf1}}
\qquad
\subfloat[After assembly]{
\includegraphics[scale=0.25]{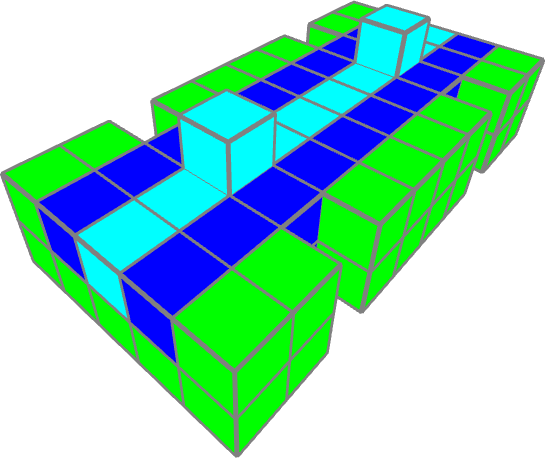}
\label{fig5_28:sf2}}
\caption{3D structure build out of 2-by-n structures.}
\label{fig5_28:fig}
\end{figure}

With the above structures one might still be a little skeptical, so let us try to mimic the well know three dimensional brick system from a Danish toy company, however built out of our 2D shapes (Figure \ref{fig5_28:fig}).
From these little sketches it should become obvious that we can build long rods, ladders or tracks, two dimensional lattices, walls, and basically any three dimensional structures.

\section{Simple Machines}

After these preliminary sketches of a somewhat handwaving nature, the question that poses itself is, can we actually build machines that perform the tasks of matching, copying and building?  To get some feeling of what can be accomplished with our blocks and their properties listed above, we will build simple machines, simulate them and visualize their workings via short, animated sequences.

\subsection{A simple Conveyor}
The simplest form of the conveyor consists of four normal blocks and two movers positioned in opposite directions all glued together.  The timing of the movers is such that first the one to the left expands and then the one to the right.  The movers will automatically contract after a little while.
\begin{figure}[h]
\centering
\subfloat[Step 1]{
\includegraphics[scale=0.25]{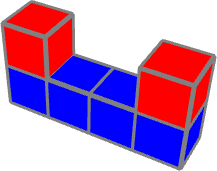}
\label{fig4_12:sf1}}
\qquad
\subfloat[Step 2]{
\includegraphics[scale=0.25]{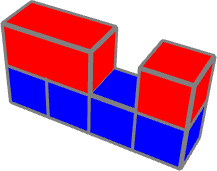}
\label{fig4_12:sf2}}
\qquad
\subfloat[Step 3]{
\includegraphics[scale=0.25]{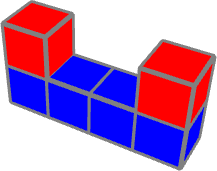}
\label{fig4_12:sf3}}
\qquad
\subfloat[Step 4]{
\includegraphics[scale=0.25]{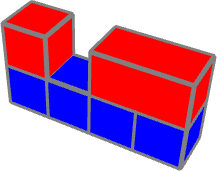}
\label{fig4_12:sf4}}
\qquad
\subfloat[Step 5]{
\includegraphics[scale=0.25]{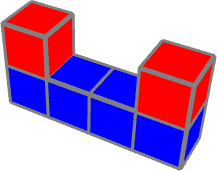}
\label{fig4_12:sf5}}
\caption{Simple Conveyor}
\label{fig4_12:fig}
\end{figure}
If we place something between the two movers, it will move first to the right, and then to the left, back and forth.  Turn it 180 degrees and it will move in the opposite direction.

For most of our applications, we will need the conveyor to move something by two units of length.  To accomplish this, we can expand the simple conveyor above and add a few additional movers.
\begin{table}[h!]\small
  \begin{center}
  \begin{tabular}{cccccc}
\includegraphics[scale=0.15]{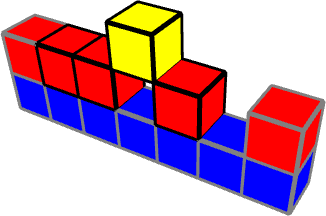} & \includegraphics[scale=0.15]{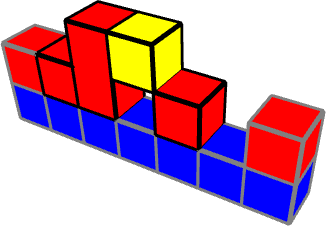} & \includegraphics[scale=0.15]{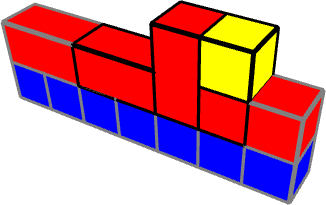} & \includegraphics[scale=0.15]{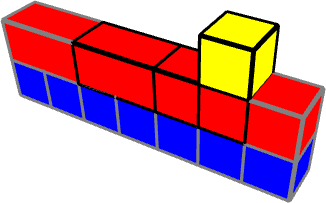} & \includegraphics[scale=0.15]{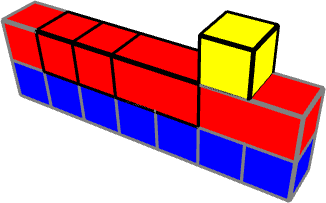} & \includegraphics[scale=0.15]{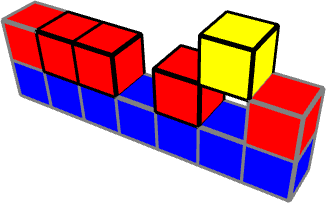}   \\
Step 1 & Step 2 & Step 3 & Step 4 & Step 5 & Step 6 \\
  \end{tabular}
\end{center}
\end{table}
It should be obvious that we can generalize this to a move-by-n conveyor.  Notice, we assume some kind of weak binding force, like a van-der-Waals type force, for the additional movers to be able to move, but still stay attached.

\begin{figure}[h]
\centering
\includegraphics[scale=0.25]{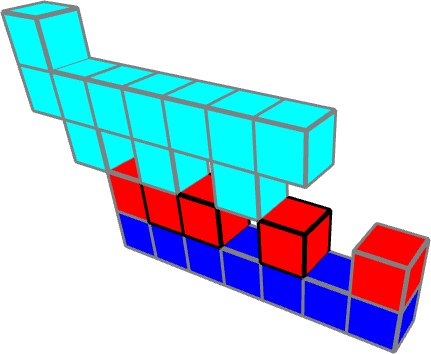}
\caption{Conveyor belt}
\label{fig4_15:fig}
\end{figure}

For moving a single block, the above might seem as overkill.  But as soon as we consider the following example (Figure \ref{fig4_15:fig}), it should be obvious why we call it the conveyor.
One way to think of the conveyor is as a conveyor, where the conveyor moves other stuff.  However, using the principle of relative motion, it should be clear that the conveyor can also move itself.  It could also move itself along tracks, kind of like kinesin, or could provide the motility function of a flagellum.

From a programming language perspective, the conveyor corresponds to a loop structure.

\subsection{A simple Matcher}
The matcher is a central component of several of our machines.  Let's first look at the matcher by itself.  It consists of a frame (yellow) and a mover (red).  
If we place a codon (blue) in the lower front part and anticodon (light blue) in the upper back part, two things can happen: codon and anticodon match, 
or codon and anticodon do not match.
\begin{figure}[h]
\centering
\subfloat[Step 1]{
\includegraphics[scale=0.25]{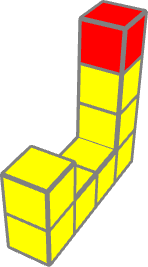}
\label{fig4_16:sf1}}
\qquad
\subfloat[Step 2]{
\includegraphics[scale=0.25]{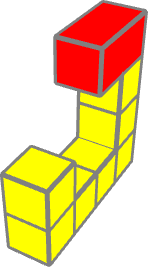}
\label{fig4_16:sf2}}
\qquad
\subfloat[Step 3]{
\includegraphics[scale=0.25]{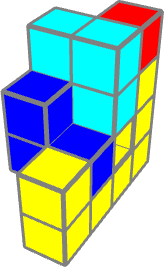}
\label{fig4_16:sf3}}
\qquad
\subfloat[Step 4]{
\includegraphics[scale=0.25]{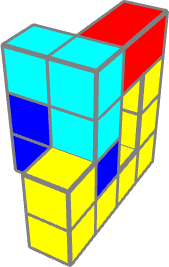}
\label{fig4_16:sf4}}
\qquad
\subfloat[Step 5]{
\includegraphics[scale=0.25]{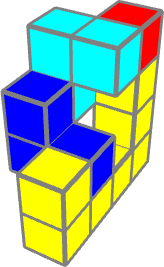}
\label{fig4_16:sf5}}
\qquad
\subfloat[Step 6]{
\includegraphics[scale=0.25]{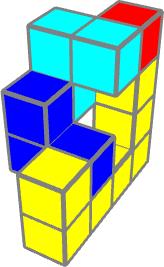}
\label{fig4_16:sf6}}
\caption{Simple matcher}
\label{fig4_16:fig}
\end{figure}
If they do not match, the anticodon will just remain where it was.  If we combine this with a conveyor that moves anticodons from left to right, we see that this will move only the anticodons that match the given codon.  This can be used to sort codons and anticodons, to pick the correct tRNA, or to copy a given RNA.  Notice, that the mover is not strong enough to break the chemical bond of the frame.

From a programming language perspective, the matcher corresponds to an if-else construct.

\subsection{A simple Builder}
\begin{figure}[h]
\centering
\subfloat[Step 1]{
\includegraphics[scale=0.25]{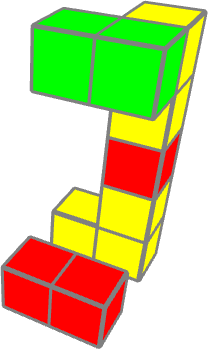}
\label{4_17:sf1}}
\qquad
\subfloat[Step 2]{
\includegraphics[scale=0.25]{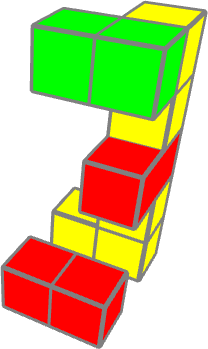}
\label{4_17:sf2}}
\qquad
\subfloat[Step 3]{
\includegraphics[scale=0.25]{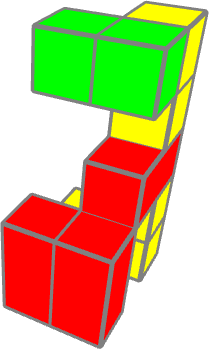}
\label{4_17:sf3}}
\qquad
\subfloat[Step 4]{
\includegraphics[scale=0.25]{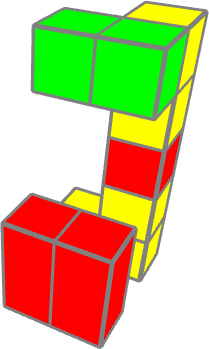}
\label{4_17:sf4}}
\qquad
\subfloat[Step 5]{
\includegraphics[scale=0.25]{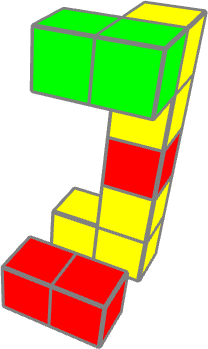}
\label{4_17:sf5}}
\qquad
\caption{Simple builder}
\label{fig4_17:fig}
\end{figure}

Recall the building process we sketched above.  We assume tRNA has been matched to mRNA and is in the right order.  The builder can build two-dimensional structures that are two blocks wide and an arbitrary number of blocks high.  We first consider the builder just by itself (Figure \ref{fig4_17:fig}).  It consists of some support structure (yellow), some movers (red) and some glue blocks (green).

\begin{figure}[h!]
\centering
\includegraphics[scale=0.25]{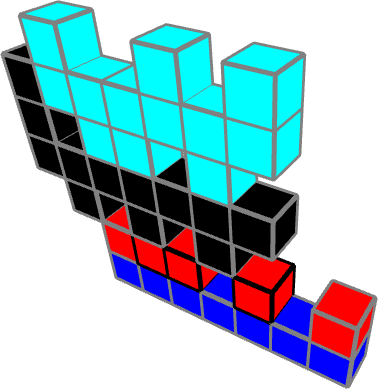}
\caption{Conveyor belt carrying tRNA.}
\label{fig4_18:fig}
\end{figure}

To see the builder in action, let us recall the conveyor belt, now filled with tRNAs already in the correct order (Figure \ref{fig4_18:fig}).

To understand how the builder works, we will go through its operation in detail (Figure \ref{fig4_19:fig}).  In step 2, the conveyor moves by two to the right.  Step 3, moves the foremost tRNA to the front.
\begin{figure}[h]
\centering
\subfloat[Step 1]{
\includegraphics[scale=0.25]{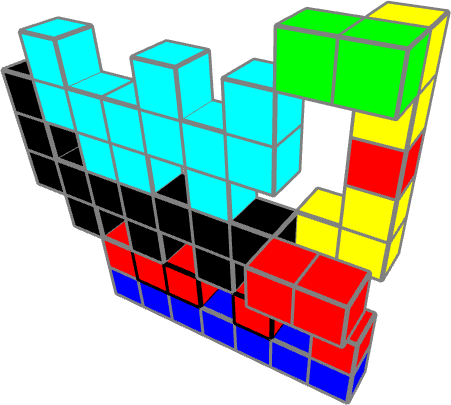}
\label{fig4_19:sf1}}
\qquad
\subfloat[Step 2]{
\includegraphics[scale=0.25]{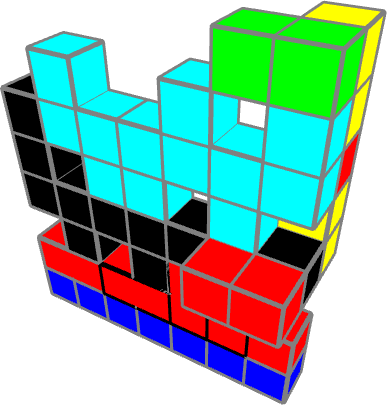}
\label{fig4_19:sf2}}
\qquad
\subfloat[Step 3]{
\includegraphics[scale=0.25]{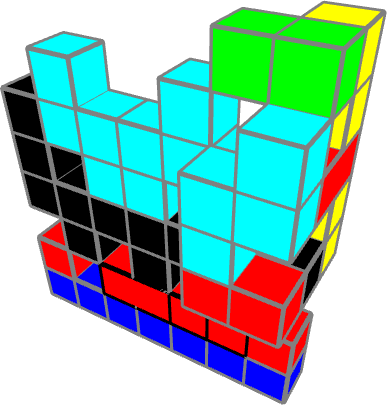}
\label{fig4_19:sf3}}
\end{figure}
\setcounter{figure}{11} 
\setcounter{subfigure}{3}
In step 4 the first tRNA is moved up by one, where it now is next to the gluer, which unglues the uppermost block from the tRNA (yellow now).  In step 5 the conveyor moves again two to the right, and step 6 moves the foremost tRNA off the conveyor.
\begin{figure}[h]
\centering
\subfloat[Step 4]{
\includegraphics[scale=0.25]{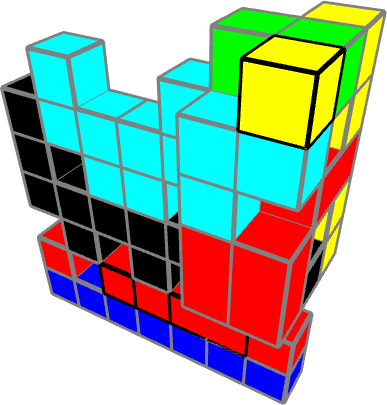}
\label{fig4_19:sf4}}
\qquad
\subfloat[Step 5]{
\includegraphics[scale=0.25]{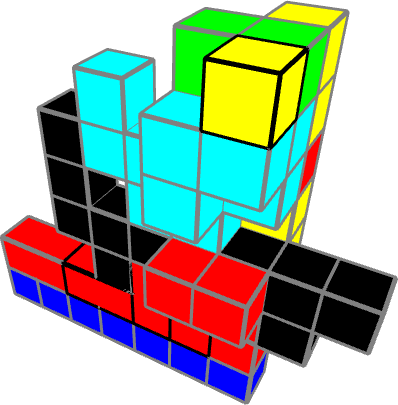}
\label{fig4_19:sf5}}
\qquad
\subfloat[Step 6]{
\includegraphics[scale=0.25]{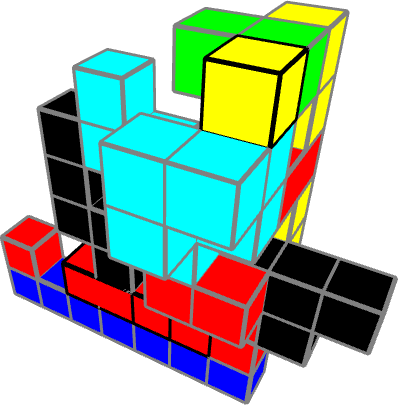}
\label{fig4_19:sf6}}
\end{figure}
\setcounter{figure}{11} 
\setcounter{subfigure}{6}
Step 7 moves the second tRNA up, unglues the uppermost block from the tRNA, and glues it to the block left over from the first tRNA (dark grey).  In step 8 the conveyor moves again two to the right, and step 9 moves the foremost tRNA off the conveyor.
\begin{figure}[h]
\centering
\subfloat[Step 7]{
\includegraphics[scale=0.25]{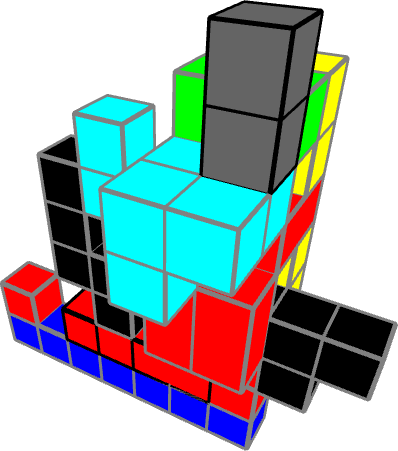}
\label{fig4_19:sf7}}
\qquad
\subfloat[Step 8]{
\includegraphics[scale=0.25]{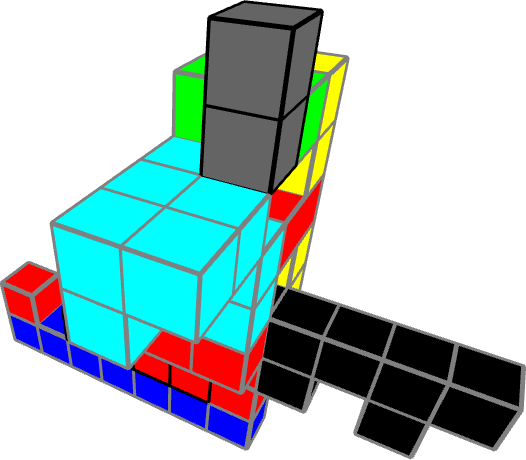}
\label{fig4_19:sf8}}
\qquad
\subfloat[Step 9]{
\includegraphics[scale=0.25]{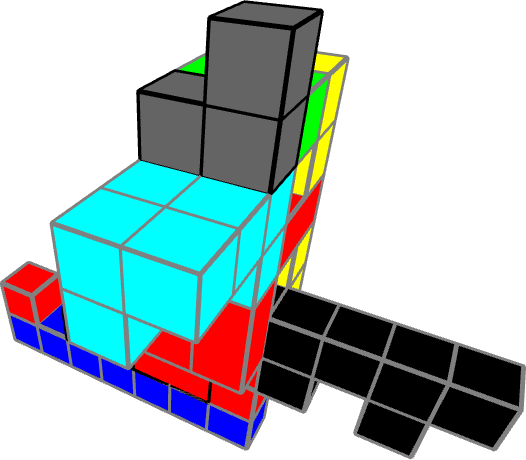}
\label{fig4_19:sf9}}
\caption{Simple builder in action.}
\label{fig4_19:fig}
\end{figure}
Finally, in step 10 the last tRNA is moved up, disconnected and glued to the pieces from the other two tRNAs.  What we see is the final product in dark grey.

You may notice two things: the emergence of a two-dimensional structure in dark grey, and the left over tRNA blocks are moved away, where they can be recycled.  With this we can build any 2-by-n two-dimensional structure.  The shape of the emerging structure solely depends on the order of the tRNA coming in.

\subsection{Other Machines}
We can build the tRNA from scratch, and initially this is also necessary.  But constructing codons and anticodons from scratch each time is a wasteful process.  For instance, to make one tRNA, we need four tRNA's in the 2-tRNA scenario and seven tRNA's in the 4-tRNA scenario.  We also would first have to shredder old tRNA and then we need four cycles to rebuild the tRNA from scratch.  If instead we simply add one piece, the piece we just lost in the build process, recycling is a very simple procedure.
In general, we want to reuse our codons, anticodons and tRNA as much as possible.  

The recycler needs to be supplied by the right types of blocks.  For this a sorter is essential.
The sorter is a very crucial component for the recycler, but in general, we need to be able to distinguish between the different block types.  If our different blocks would be of different sizes, then sorting them is a straightforward process.  If, however, blocks can not be distinguished by their size or other physical property, then things get more complicated.  In addition, the sorter also kind of needs to know the direction of mover and gluer.  For this study, we ignore these details.  But it should be clear, that this is an important, non-trivial issue that needs to be solved.
Also, surprisingly difficult is the shredder.  If we want to reuse our building blocks, then we need to be able to take them apart again.  This can be done by giving additional functionality to the gluer or by introducing an entirely new block type.  Although at first thought, the shredder may seem to be not very important, I believe it actually is.  First of all, our building blocks, especially the movers and gluers, are far to precious to just throw them on a garbage pile.  But second, consider the beginning of evolution: if evolution is started through random processes, a vast amount of garbage is produced in this trial and error stage of evolution.

One could think of many more little machines performing in general helpful tasks.  One's that come immediately to mind are collector, filter, and scaffolding.  Most likely, in the beginning the density of blocks might be rather low.  In this case a collector might be useful, that collects blocks, and increases the density of useful material.  The filter could be useful when sorting out garbage from useful material, maybe even partially constructed material.

\section{The Constructor}

Although the simple builder just shown is able to construct any 2-by-n two-dimensional structure, it can not build itself, hence it is not universal.  Instead of trying to construct a machine that can construct itself, we will construct two machines: a builder and an assembler.  All our machines will consist of 2-by-n two-dimensional components, which the builder can build.  And all of our machines can be assembled out of these components, which is what the assembler does.  If we can construct both the builder and the assembler this way, we have accomplished our goal.

If the builder can build all the components, including components for itself and the assembler, and the assembler can assemble all machines, including the builder and itself, then the combination of builder and assembler performs basically the same task as von Neumann's universal constructor.

\subsection{4-tRNA}
Looking at the machines we just built, we learned that we need at least three different kinds of building blocks, and if we want to be able to build 2-by-n structures we need to be able to place normal blocks in two different positions.  This leads to the requirement that we need at least four different tRNA for encoding.  As we have seen, there are six different 4-codons, hence 4-codons should suffice.  We therefore define the following 4-tRNA, keeping the names of the anticodons.  Notice, the last two are still available.
\begin{table}[h!]\small
  \begin{center}
  \begin{tabular}{cccccc}
\includegraphics[scale=0.25]{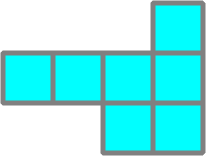} & \includegraphics[scale=0.25]{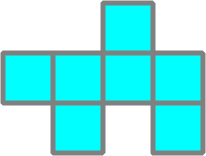} & \includegraphics[scale=0.25]{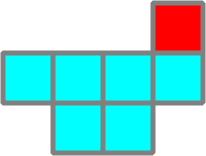} & \includegraphics[scale=0.25]{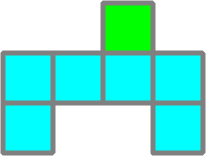} & \includegraphics[scale=0.25]{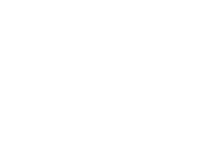} & \includegraphics[scale=0.25]{figures/4_4-tRNA_f.png}   \\
A: Normal right & B: Normal left & C: Mover & D: Gluer & E: Still available & F: Still available \\
  \end{tabular}
\end{center}
\end{table}

\subsection{The Builder}
Compared to the simple version of our builder, we need to add an additional mover at the bottom, because it needs to move 4-anticodons now, and the conveyor belt has to shift by four each time, instead of two.
\begin{figure}[h]
\centering
\includegraphics[scale=0.25]{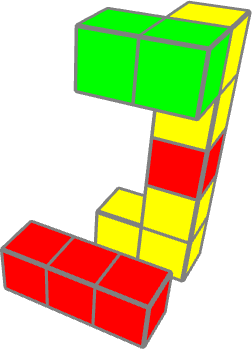}
\caption{The Builder}
\label{fig5_24:fig}
\end{figure}
It should be noted, that we have only given the bare minimum structure for the builder to make it easy to grasp its workings.  It should be clear, that the bottom movers can not just "levitate", instead they should somehow be connected to the rest of the builder construction.  We will discuss this in more detail below.

\subsection{Components}
If we look at our builder we realize, that it is wider than a 2-by-n structure.  Therefore, we can not hope to build the builder in one single piece.  But instead, we can split our machines up into 2-by-n structures, which we call components.  We will show that our builder can build all the basic components for all our machines. 

\subsubsection{Conveyor Components}
Recall our simple conveyor, Figure \ref{fig4_12:fig}.  Hypothetically, we could construct the simple conveyor directly, but the subtle problem is that the two mover pieces point in different directions.  However, our 4-tRNA only points in one given direction.  Hence, we will not attempt to construct the simple conveyor in one piece, but instead in three pieces.  The following three simple pieces can be constructed with our builder and the above 4-tRNA.  With these components, we can assemble the simple conveyor.
\begin{table}[htbp]\small
  \caption{Conveyor components}
  \label{tab5_1:tab}
  \begin{center}
  \begin{tabular}{|c|c|c|}
\hline
Component & BDL \\\hline
\includegraphics[scale=0.25]{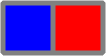} & AC \\\hline
\includegraphics[scale=0.25]{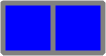} & 2x AA \\\hline
  \end{tabular}
\end{center}
\end{table}

It should be noted, that extending our codons, for instance using the unused F codon, we could build the conveyor in one piece.  But that would still leave us with the timing issue: we still need to explicitly encode the timing in our tRNA.  We could do this by using 5-codons or 6-codons, but we will not do this here.

\subsubsection{Builder Components}
\begin{figure}[h]
\centering
\includegraphics[scale=0.25]{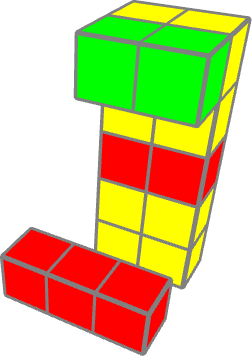}
\caption{Builder}
\label{fig5_26:fig}
\end{figure}

To show that our builder can also be build out of components, we first make a small modification to it (Figure \ref{fig5_26:fig}).  
With this change, only two kinds of components need to be build (table \ref{tab5_2:tab}).  Hence, the builder can build its own components.
\begin{table}[htbp]\small
  \caption{Builder components}
  \label{tab5_2:tab}
  \begin{center}
  \begin{tabular}{|c|c|c|}
\hline
Component & BDL \\\hline
\includegraphics[scale=0.25]{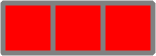} & CCC \\\hline
\includegraphics[scale=0.25]{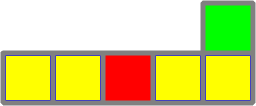} & DAACAA \\\hline
  \end{tabular}
\end{center}
\end{table}

\subsection{Builder Description Language (BDL)}
We have seen it before, but let us be a little more explicit about the builder description language.  It is the language the builder understands, and it is equivalent to von Neumann's description function,  $\Phi(X)$.  We call it our "genetic code".  For instance, the genetic code to build the builder is "CCC\_DAACAA\_DAACAA" and for the simple conveyor it is "AC\_AC\_AA", where the underline indicates something like a stop codon and the letters refer to the 4-tRNA.

\subsection{The Assembler}
\begin{figure}[h]
\centering
\includegraphics[scale=0.25]{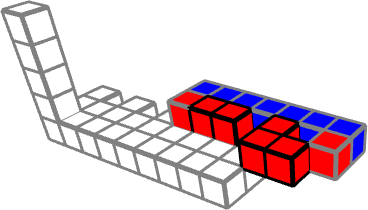}
\caption{Assembler with transport structure.}
\label{fig5b_27b:fig}
\end{figure}

The assembler needs to assemble the components into machines.  This sounds complicated, but the assembler is surprisingly simple: it consists of a move-by-two conveyor, two simple movers and a transport structure (Figure \ref{fig5b_27b:fig}).

\subsubsection{Assembler Components}
Before showing the assembler in action, we need to make sure the builder can build the components of the assembler.
\begin{table}[htbp]\small
  \caption{Assembler components}
  \label{tab5_3:tab}
  \begin{center}
  \begin{tabular}{|c|c|c|}
\hline
Component & BDL \\\hline
\includegraphics[scale=0.25]{figures/4_builder_1_3.png} & CCA \\\hline
\includegraphics[scale=0.25]{figures/4_builder_1_4.png} & FAAAAAAF \\\hline
  \end{tabular}
\end{center}
\end{table}

\subsubsection{Assembler assembles Builder}
To see how the assembler works, we will consider first the assembly of the builder.  The components of the builder have been build by the builder, as indicated above.  Those components are in the correct order on the transporter (Figure \ref{fig5b_9:fig}).  Notice, that we have "filler" pieces in between.  The assembly process is as follows:  In a first step, the transporter is moved to the right by two.  Then the two movers move two components off of the transporter.  These steps are repeated three times as shown in Figure \ref{fig5b_9:fig}.
\begin{figure}[h!]
\centering
\subfloat[Step 1]{
\includegraphics[scale=0.25]{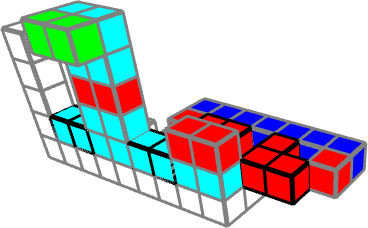}
\label{fig5b_9:sf1}}
\qquad
\subfloat[Step 2]{
\includegraphics[scale=0.25]{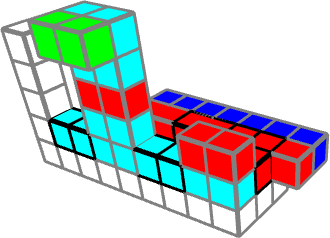}
\label{fig5b_9:sf2}}
\subfloat[Step 3]{
\includegraphics[scale=0.25]{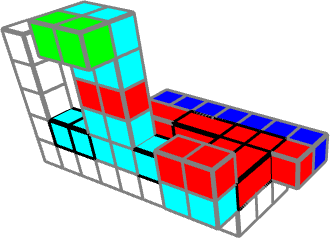}
\label{fig5b_9:sf3}}
\qquad
\subfloat[Step 4]{
\includegraphics[scale=0.25]{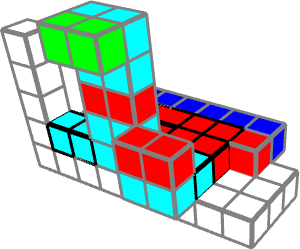}
\label{fig5b_9:sf4}}
\end{figure}

\setcounter{figure}{15} 
\setcounter{subfigure}{4}

\begin{figure}[h!]
\centering
\subfloat[Step 5]{
\includegraphics[scale=0.25]{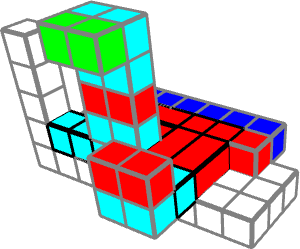}
\label{fig5b_9:sf5}}
\qquad
\subfloat[Step 6]{
\includegraphics[scale=0.25]{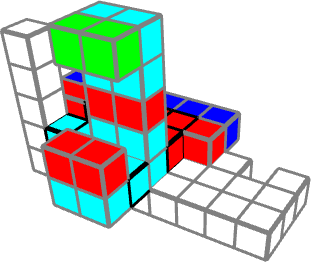}
\label{fig5b_9:sf6}}
\subfloat[Step 7]{
\includegraphics[scale=0.25]{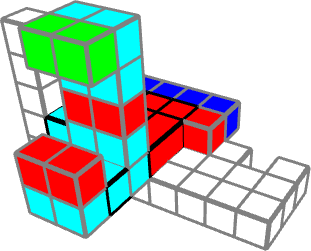}
\label{fig5b_9:sf7}}
\qquad
\subfloat[Step 8]{
\includegraphics[scale=0.25]{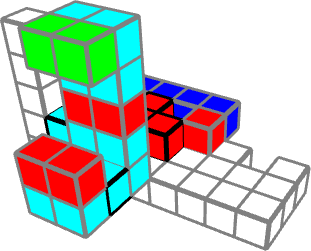}
\label{fig5b_9:sf8}}
\caption{Assembler assembles builder.}
\label{fig5b_9:fig}
\end{figure}

At the end of the procedure, step 8, we recognize our builder.  If the builder is given the instruction  "CA\_CA\_A\_A\_DAACAA\_DAACAA" it will produce the components for the builder in the right order.

\subsubsection{Assembler assembles Assembler}
\begin{figure}[h!]
\centering
\subfloat[Step 1]{
\includegraphics[scale=0.25]{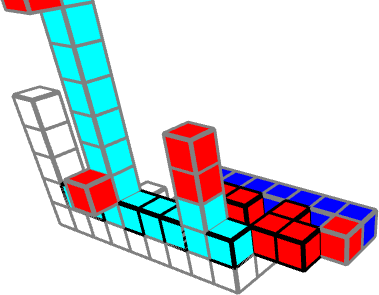}
\label{fig5b_10:sf1}}
\qquad
\subfloat[Step 2]{
\includegraphics[scale=0.25]{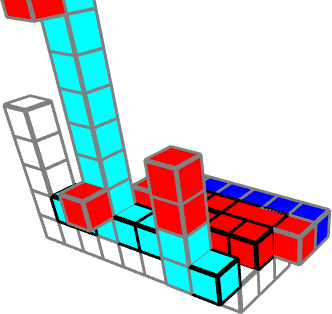}
\label{fig5b_10:sf2}}
\subfloat[Step 3]{
\includegraphics[scale=0.25]{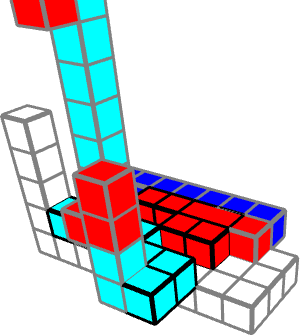}
\label{fig5b_10:sf3}}
\qquad
\subfloat[Step 4]{
\includegraphics[scale=0.25]{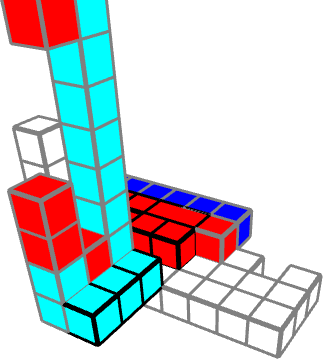}
\label{fig5b_10:sf4}}
\caption{Assembler assembles assembler.}
\label{fig5b_10:fig}
\end{figure}

Finally, we demonstrate that the assembler can assemble itself, Figure \ref{fig5b_10:fig}.

On the transporter are the components needed to build the assembler.  After going through the same process as above, we notice that in step 4 we have a finished assembler.  If the builder is given the instruction "A\_CCAA\_A\_A\_A\_FAAAAAAAFA" it will produce the components for the assembler in the right order, where we used a modified F-tRNA for the movers that are one to the left.

The careful reader will notice a few omissions.  We have neglected to show that during assembly, the components need to be glued together.  This could easily be done by adding a gluer.  Also, the transporter is not part of the assembler, and hence the assembly process.  And, maybe more critical, the freely moving movers are not part of the assembly procedure.

\section{The Copier}

The copier is a little more complicated than the builder.  We will only show a 2-copier, which copies RNA made up of  2-codons, but the generalization should be obvious.  First, consider a strand of RNA that we want to copy, Figure \ref{fig6_29:fig}:
\begin{figure}[h!]
\centering
\subfloat[RNA with encasing]{
\includegraphics[scale=0.25]{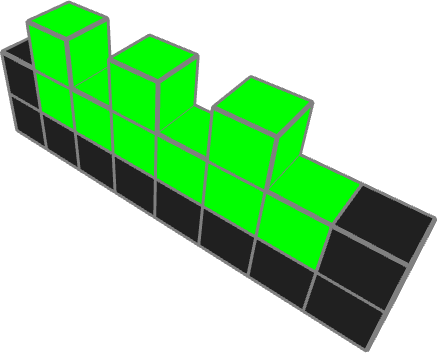}
\label{fig6_29:sf1}}
\qquad
\subfloat[Conveyor belt with anticodons]{
\includegraphics[scale=0.25]{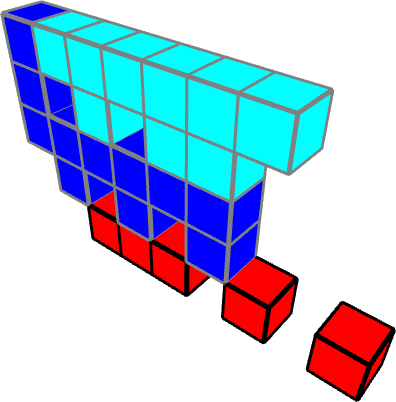}
\label{fig6_29:sf2}}
\caption{RNA and conveyor belt.}
\label{fig6_29:fig}
\end{figure}
green is the RNA and black is the encasing.  To understand the copier we must recall the conveyor belt from above, but instead of carrying tRNA, it will carry anticodons.  The arrangement of the anticodons is random.
Comparing the RNA codons and the anticodons from the conveyor, Figure \ref{fig6_29:fig}, notice that the first anticodon does not match, but the other two anticodons will match.  Also, recall the matcher (Figure \ref{fig4_16:fig}).

\begin{figure}[h]
\centering
\subfloat[Copier front]{
\includegraphics[scale=0.25]{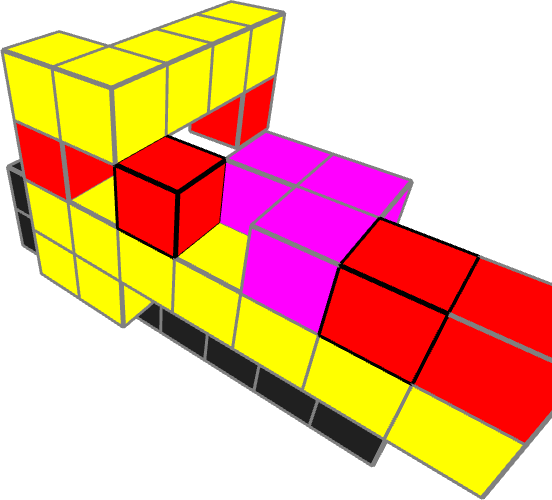}
\label{fig6_32:sf1}}
\qquad
\subfloat[Copier back]{
\includegraphics[scale=0.25]{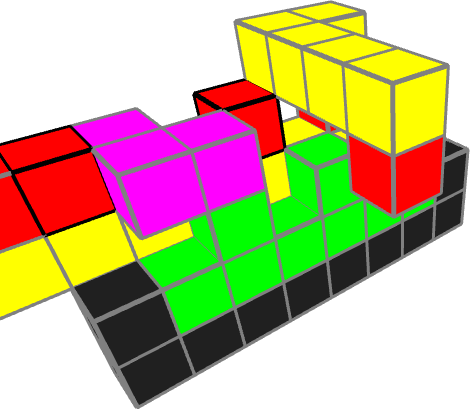}
\label{fig6_32:sf2}}
\caption{Copier}
\label{fig6_32:fig}
\end{figure}

With this in mind, let us consider the empty copier (Figure \ref{fig6_32:fig}), left is a look from the front, right is a look from the back.
In the back you should recognize the matcher part.  In the front is a 2-mover that moves the RNA strand by two to the left via the magenta ledger.  However, it will do this move only, if there has been a match between codon of the RNA and anticodon.  Without anticodons it will not move at all.

The little animation sequence shows how the copier works.  The tRNA conveyor moves by two to the left.  In step 2, the matcher pushes the anticodon, which happens to match, onto the RNA strand.  In step 3, the ledger catches the tRNA and moves the whole RNA strand to the right.  Step 4 shows the copier after one iteration: the tRNA conveyor has moved to the left, the RNA to the right.  To see that the RNA comes out in the back, we have shown step 5, which is the result after one more match, and clearly, the RNA strand comes out in the back, carrying a (negative) copy.
\begin{table}[h!]\small
  \begin{center}
  \begin{tabular}{ccccc}
\includegraphics[scale=0.15]{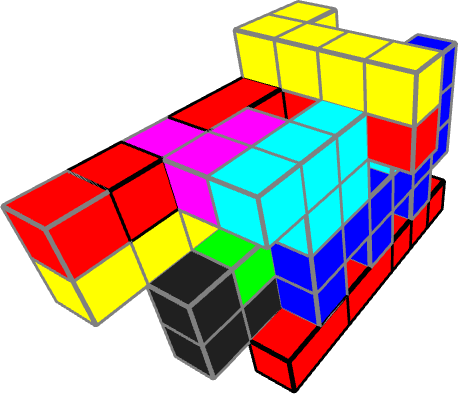} &
\includegraphics[scale=0.15]{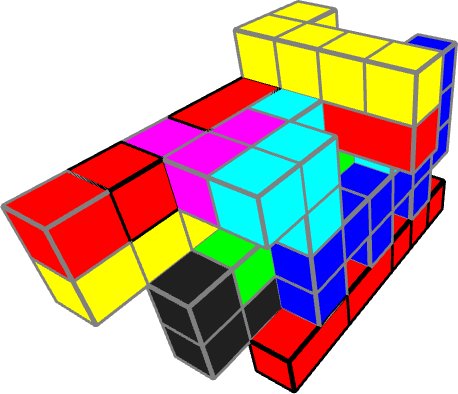} &
\includegraphics[scale=0.15]{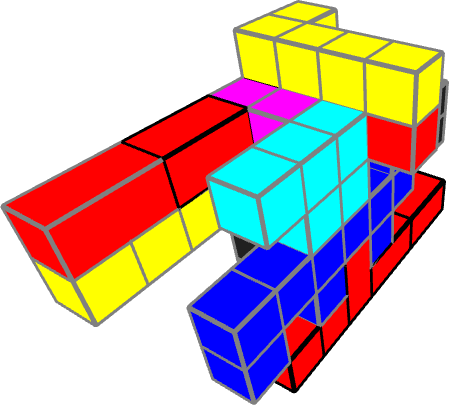} &
\includegraphics[scale=0.15]{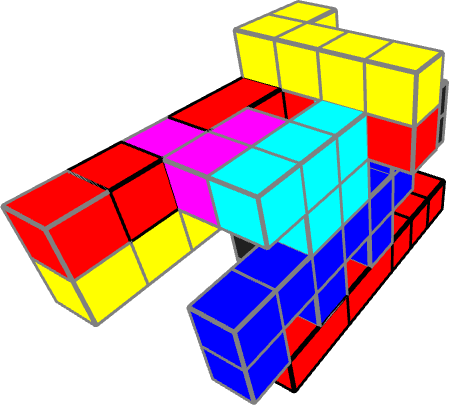} &
\includegraphics[scale=0.14]{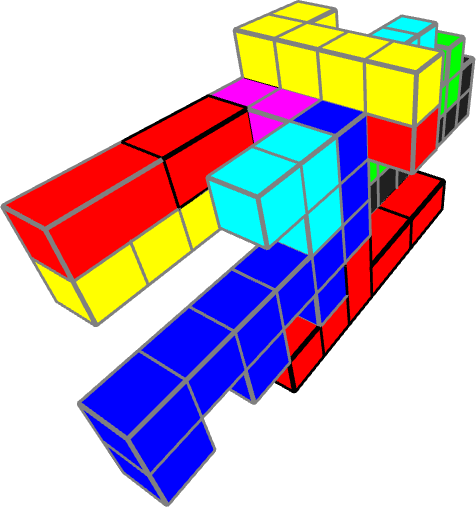}    \\
Step 1 & Step 2 & Step 3 & Step 4 & Step 5 \\
  \end{tabular}
\end{center}
\end{table}

\pagebreak

\subsection{Components}
We need to show, that we can build the copier with our builder.  The following components, when assembled correctly will form a copier, and each of these components can be build by our builder.
\begin{table}[htbp]\small
  \caption{Copier components}
  \label{tab6_3:tab}
  \begin{center}
  \begin{tabular}{|c|c|c|}
\hline
Component & BDL \\\hline
\includegraphics[scale=0.25]{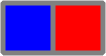} & AC \\\hline
\includegraphics[scale=0.25]{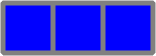} & AAA \\\hline
\includegraphics[scale=0.25]{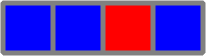} & AACA \\\hline
\includegraphics[scale=0.25]{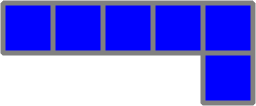} & AAAAAB \\\hline
  \end{tabular}
\end{center}
\end{table}
Hence, the genetic code for our copier is "AC\_AC\_AAA\_AACA\_AAAAAB".

As you may have noticed, our copier does not produce a copy, but rather a negative copy.  This is no problem, because we simply run it twice, and then we will have a copy.  Also you notice, that we copy only 2-codon RNA.  But the mechanism is easily generalized to 4-codons.
It is not difficult to see that only a slight modification to the copier is needed for translation to happen, that is turning mRNA into tRNA.  The conveyor will carry tRNA instead of anticodons.  In the copier above, the arm needs to be one block size higher, for the tRNA to fit.
What we have not shown is how to assemble the copier.

\pagebreak

\section{Turing Machines}

\begin{figure}[h]
\centering
\includegraphics[scale=0.5]{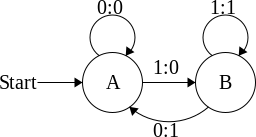}
\caption{Times two}
\label{fig7_32:fig}
\end{figure}

It is very interesting to see that no additional ingredients are needed to implement Turing machines \cite{turing1936computable,minsky1967computation} with our construction.  Turing machines are effectively described by their state-diagram, for instance Firgure \ref{fig7_32:fig} shows the state diagram of a binary multiply by two Turing machine.
It consists of two states, a left-moving and a right-moving state, with transitions indicated by the arrows.  To build a Turing machine one needs a tape, a head that can move to the left and right, and that can read and write.  It has a fixed number of symbols and a fixed number of internal states. Clearly the tape is our RNA, and our codons are the symbols.  Reading is done by our matcher.  What we need to worry about is writing, moving in two directions, and keeping internal state.  We want our Turing machine to be as close to the above mechanisms as possible, but it is also possible to construct Turing machines based on totally different designs.

\subsection{1-Complement}
Let us consider one of the simplest Turing machines, the binary one-complement.  It has only one state,
\begin{figure}[h]
\centering
\includegraphics[scale=0.5]{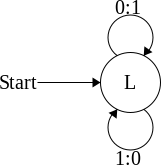}
\caption{1-complement}
\label{fig7_33:fig}
\end{figure}
and for simplicity, we do not worry about halting.  If for instance the input is "0110100" than the output would be "1001011".

We represent the input numbers by the following 2-codons.
\begin{table}[h!]\small
  \begin{center}
  \begin{tabular}{cc}
\includegraphics[scale=0.25]{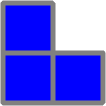} &
\includegraphics[scale=0.25]{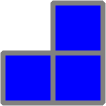} \\
0 & 1 \\
  \end{tabular}
\end{center}
\end{table}
Next, we match them with their respective anticodons.  But instead of having simple blocks as their payload, they now have 2-codons as their payload.
\begin{table}[h!]\small
  \begin{center}
  \begin{tabular}{cc}
\includegraphics[scale=0.25]{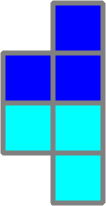} &
\includegraphics[scale=0.25]{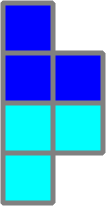} \\
tRNA 0 & tRNA 1 \\
  \end{tabular}
\end{center}
\end{table}

For our sample input above, our tape will look like this:
\begin{table}[h!]\small
  \begin{center}
  \begin{tabular}{ccccccc}
\includegraphics[scale=0.25]{figures/tm_2-codons_a.png} &
\includegraphics[scale=0.25]{figures/tm_2-codons_b.png} &
\includegraphics[scale=0.25]{figures/tm_2-codons_b.png} &
\includegraphics[scale=0.25]{figures/tm_2-codons_a.png} &
\includegraphics[scale=0.25]{figures/tm_2-codons_b.png} &
\includegraphics[scale=0.25]{figures/tm_2-codons_a.png} &
\includegraphics[scale=0.25]{figures/tm_2-codons_a.png} \\
0 & 1 & 1 & 0 & 1 & 0 & 0 \\
  \end{tabular}
\end{center}
\end{table}

Which will be matched by the following tRNAs:
\begin{table}[h!]\small
  \begin{center}
  \begin{tabular}{ccccccc}
\includegraphics[scale=0.25]{figures/tm_tRNA_a.png} &
\includegraphics[scale=0.25]{figures/tm_tRNA_b.png} &
\includegraphics[scale=0.25]{figures/tm_tRNA_b.png} &
\includegraphics[scale=0.25]{figures/tm_tRNA_a.png} &
\includegraphics[scale=0.25]{figures/tm_tRNA_b.png} &
\includegraphics[scale=0.25]{figures/tm_tRNA_a.png} &
\includegraphics[scale=0.25]{figures/tm_tRNA_a.png} \\
tRNA 0 & tRNA 1 & tRNA 1 & tRNA 0 & tRNA 1 & tRNA 0 & tRNA 0 \\
  \end{tabular}
\end{center}
\end{table}

If you look at the output, the top part, you will see it is the inverted input.  Using our matcher, it should be easy now to see how we can implement such a Turing machine.  Notice, codon and anticodon do make up the tRNA, but they are not glued together.  Meaning they can relatively easily be separated again. 

\subsection{The Writer}
\begin{figure}[h]
\centering
\subfloat[Step 1]{
\includegraphics[scale=0.2]{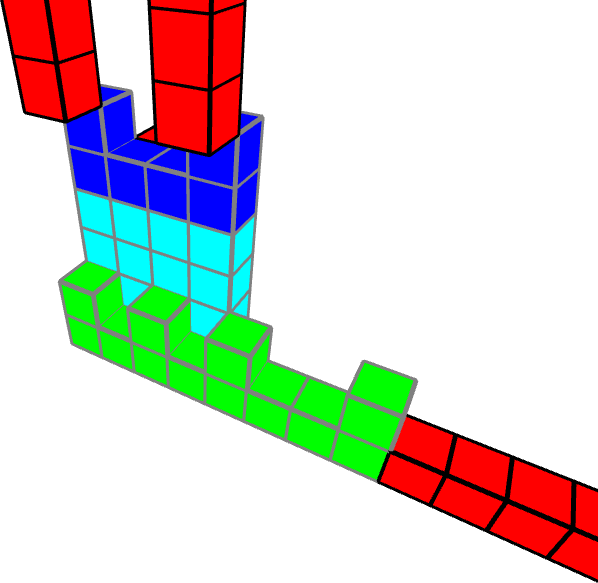}
\label{fig7_38:sf1}}
\qquad
\subfloat[Step 2]{
\includegraphics[scale=0.2]{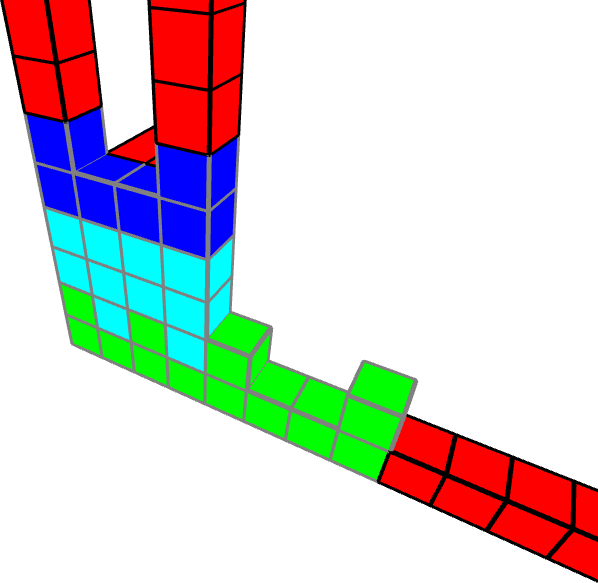}
\label{fig7_38:sf2}}
\qquad
\subfloat[Step 3]{
\includegraphics[scale=0.2]{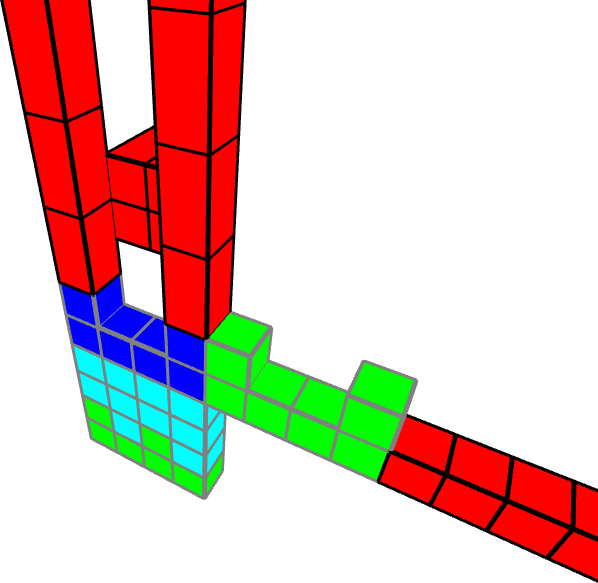}
\label{fig7_38:sf3}}
\caption{Writer}
\label{fig7_38:fig}
\end{figure}

How can the write operation be performed?  Figure \ref{fig7_38:fig} shows how this can be done.  In green we see the tape, in light blue the anticodon, and in dark blue the new codons.  In step 2 the matching is performed, and in step 3 we see the write operation: we simply push from above and replace the old, green part of the tape, with the new, blue part.  In a next step, we would shift to the next location on the tape and the whole process starts anew.

\pagebreak

\subsection{The Left-Right Conveyor}
To see how we can implement left and right movement of our tape, first consider the following small modification to our anticodons:
\begin{table}[h!]\small
  \begin{center}
  \begin{tabular}{ccc}
\includegraphics[scale=0.25]{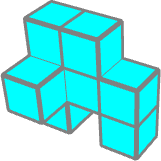} &
\includegraphics[scale=0.25]{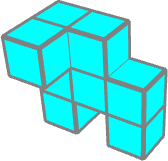} &
\includegraphics[scale=0.25]{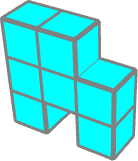}  \\
Left anticodon & Right anticodon & Halt anticodon \\
  \end{tabular}
\end{center}
\end{table}

We add a column to the left, so that our 2-codons are now three blocks wide, and we add a little peg that is extruding.  If we want a left movement, it is extruding at the lower part, if we want a right movement, it will extrude at the upper part, and if we do not want any movement, there is no peg at all.
The movement itself will be done by two conveyors, in this case move-by-3 conveyors. One for the left move and a second one, positioned up by one and rotated 180 degrees, for the right move.
\begin{figure}[h]
\centering
\subfloat[Step 1]{
\includegraphics[scale=0.25]{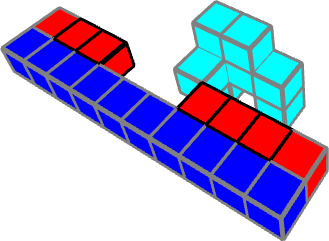}
\label{fig7_40:sf1}}
\qquad
\subfloat[Step 2]{
\includegraphics[scale=0.25]{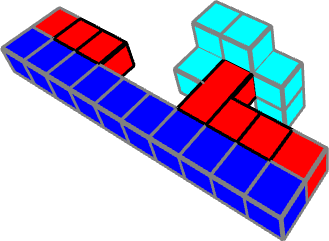}
\label{fig7_40:sf2}}
\qquad
\subfloat[Step 3]{
\includegraphics[scale=0.25]{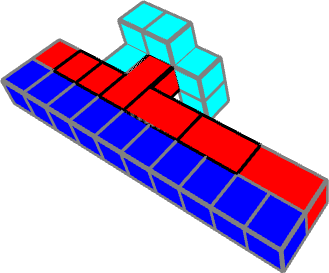}
\label{fig7_40:sf3}}
\caption{Left conveyor}
\label{fig7_40:fig}
\end{figure}



\subsection{Simple Turing Machine}
We are now able to construct a simple working Turing machine.  We have a conveyor that transports the tRNA in the back and in addition the tape (green) in front.
\begin{figure}[h]
\centering
\subfloat[Conveyor without payload.]{
\includegraphics[scale=0.25]{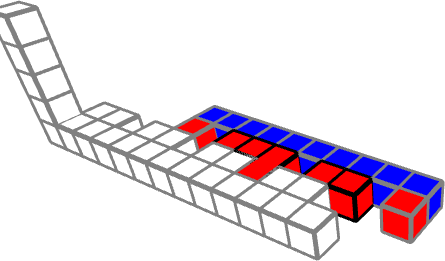}
\label{fig7_42:sf1}}
\qquad
\subfloat[Conveyor with tRNA.]{
\includegraphics[scale=0.25]{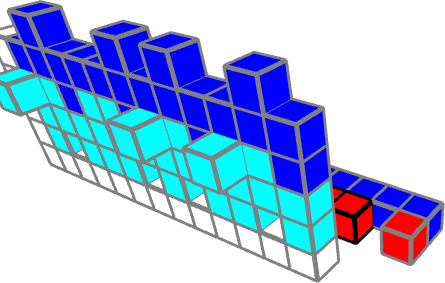}
\label{fig7_42:sf2}}
\qquad
\subfloat[Conveyor with tRNA and RNA in front.]{
\includegraphics[scale=0.25]{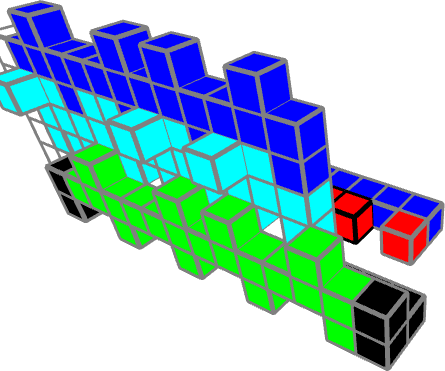}
\label{fig7_42:sf3}}
\caption{Parts of simple Turing machine.}
\label{fig7_42:fig}
\end{figure}
The basic 2-codon (green) now is one block longer, which is needed because of the left-right extension of the tRNA.  Also, the codons are inside a frame (black) that makes sure the codons stay together when moved to the left and right, and in addition, that allows for the pushing down of the write action.

\begin{figure}[h!]
\centering
\subfloat[Step 1]{
\includegraphics[scale=0.25]{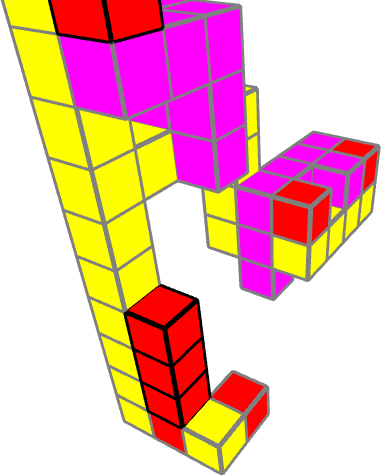}
\label{fig7_44:sf1}}
\qquad
\subfloat[Step 2]{
\includegraphics[scale=0.25]{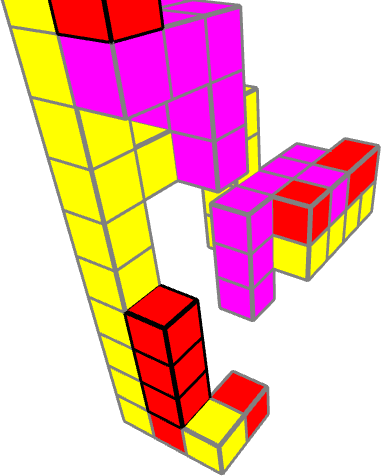}
\label{fig7_44:sf2}}
\qquad
\subfloat[Step 3]{
\includegraphics[scale=0.25]{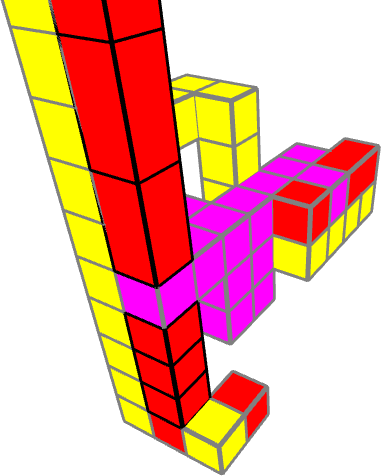}
\label{fig7_44:sf3}}
\caption{Matcher and Writer}
\label{fig7_44:fig}
\end{figure}

Figure \ref{fig7_44:fig} shows the central elements of the Turing machine.  A matcher that pushes the tRNA forward (step 2), meaning anticodon and codon, which means it has to be a little larger, but other than that the same as for the copier.  And in addition, in step 3, we perform the write operation: pushing from above like the writer.

Finally, in Figure \ref{fig7_45:fig} we can see the whole machine in action.  First, move the tRNA in front of the matcher and match.
\begin{figure}[h]
\centering
\subfloat[Step 1]{
\includegraphics[scale=0.25]{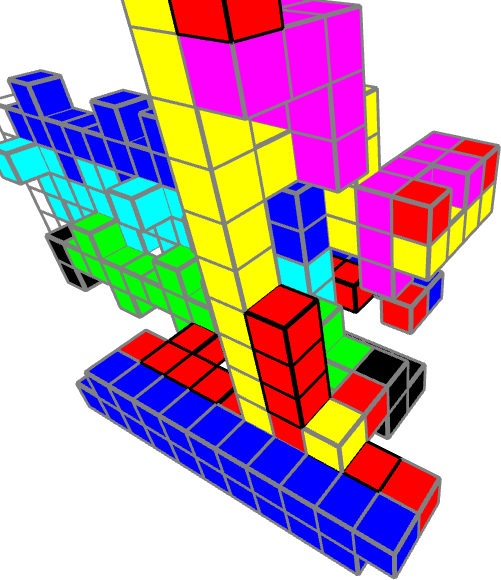}
\label{fig7_45b:sf1}}
\qquad
\subfloat[Step 2]{
\includegraphics[scale=0.25]{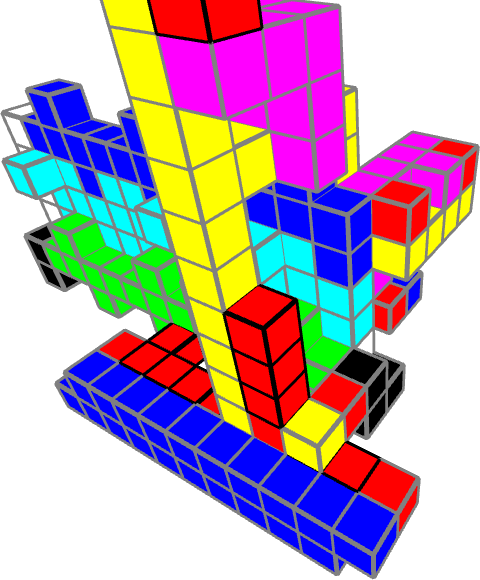}
\label{fig7_45b:sf2}}
\qquad
\subfloat[Step 3]{
\includegraphics[scale=0.25]{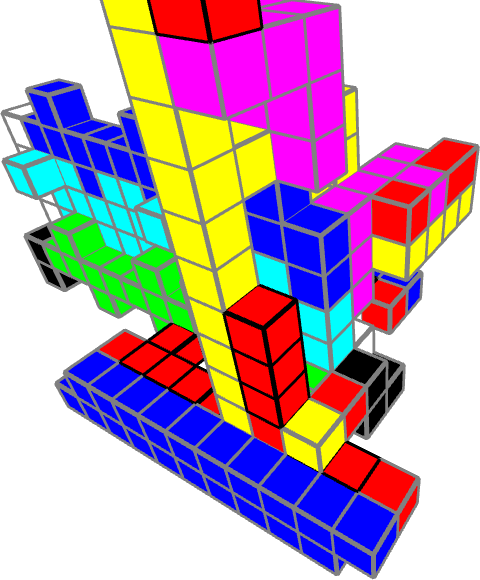}
\label{fig7_45b:sf3}}
\end{figure}
Then push down, that is perform the write operation in step 4.  In step 5, move the tape to the left or right, and step 6 moves the pusher back up, so we can start anew.
\setcounter{figure}{25} 
\setcounter{subfigure}{3}
\begin{figure}[h]
\centering
\subfloat[Step 4]{
\includegraphics[scale=0.25]{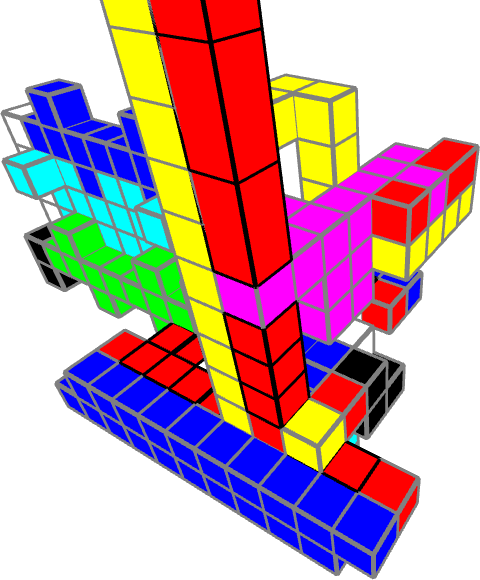}
\label{fig7_45:sf1}}
\qquad
\subfloat[Step 5]{
\includegraphics[scale=0.25]{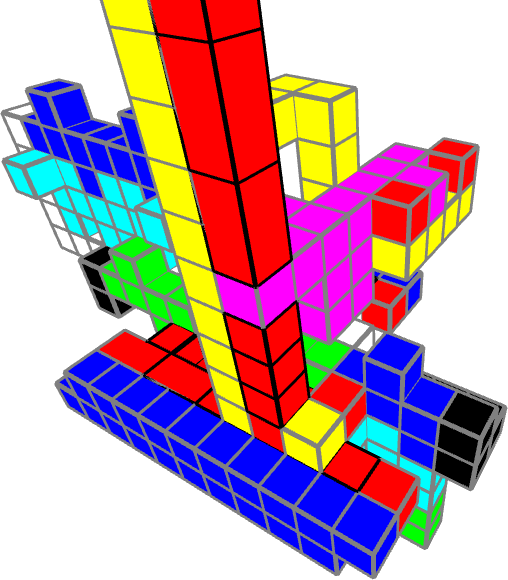}
\label{fig7_45:sf2}}
\qquad
\subfloat[Step 6]{
\includegraphics[scale=0.25]{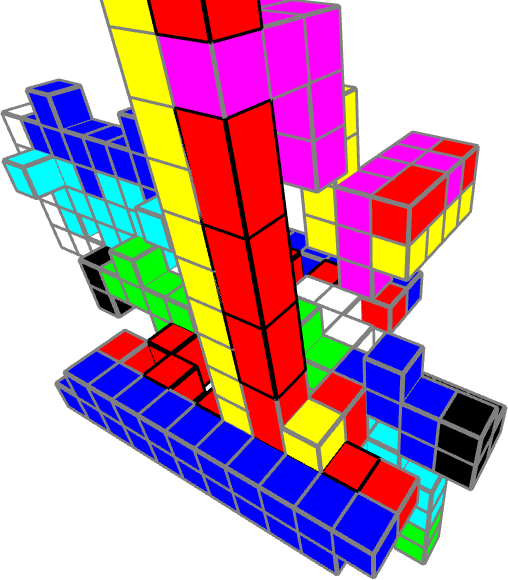}
\label{fig7_45:sf3}}
\caption{Simple Turing Machine}
\label{fig7_45:fig}
\end{figure}
This is a working Turing machine implementing the 1-complement.

\subsection{Examples}
Besides the 1-complement, the machine above can also implement at least two more Turing machines (Table \ref{tab7_4:tab}).  Let us see how they differ in their programming, that is their tRNA.
\begin{table}[htbp]\small
  \caption{Example Turing Machines}
  \label{tab7_4:tab}
  \begin{center}
  \begin{tabular}{|l|l|l|}
\hline
1-Complement & Halt & Infinite Loop \\\hline
\includegraphics[scale=0.5]{figures/tm_1_complement.png} & \includegraphics[scale=0.5]{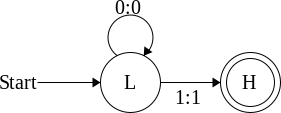}  & \includegraphics[scale=0.5]{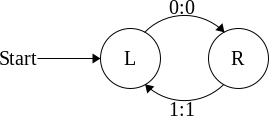} \\[20pt]
\hline
Symbols: 0, 1 &	Symbols: 0, 1 &	Symbols: 0, 1 \\\hline
States: $L$ &	States: $L, H$ &	States: $L, R$ \\\hline
Rules: & Rules: & Rules: \\
$(L, 0) \rightarrow (L, 1)$ &  $(L, 0) \rightarrow (L, 0)$ &  $(L, 0) \rightarrow (R, 0)$ \\
$(L, 1) \rightarrow (L, 0)$ &  $(L, 1) \rightarrow \; ???$    &  $(L, 1) \rightarrow \; ???$    \\
                            &  $(H, 0) \rightarrow \; ???$    &  $(R, 0) \rightarrow \; ???$    \\
                            &  $(H, 1) \rightarrow (H, 1)$ &  $(R, 1) \rightarrow (L, 1)$ \\\hline
  \end{tabular}
\end{center}
\end{table}
The rules are nothing but a lookup table: it says if you encounter the tuple (L, 0) replace it with the tuple (L, 1).  This lookup table is what our tRNA does.  To make clear of what we mean, we show the respective tRNAs for the above examples.

\pagebreak

\subsubsection{1-Complement}
First, look at the lower part of the anticodon: it encodes the 0 or 1 in the left-hand side of our rules.  The yellow block encodes the L or R.  As for the right-hand side, that is encoded in the upper part, representing the 1 or 0, respectively. 
\begin{table}[htbp]\small
  \begin{center}
  \begin{tabular}{cc}
\includegraphics[scale=0.25]{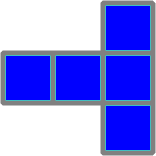} & \includegraphics[scale=0.25]{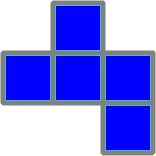}  \\
\includegraphics[scale=0.25]{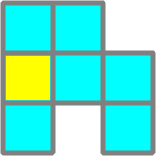} & \includegraphics[scale=0.25]{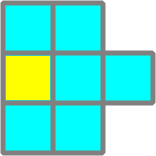}  \\
\\
(L, 0) $\rightarrow$ (L, 1) & (L, 1) $\rightarrow$ (L, 0) \\
  \end{tabular}
\end{center}
\end{table}
Notice, that in this simple Turing machine, we do not really encode the L/R part of the right-hand side.

\subsubsection{Halt}
To demonstrate the halt state, we use the lookup table below.  Again, this only works in the very limited scenario, where we start out with 0's in the input, which will trigger the move left.  As soon as it encounters a 1 in the input, it will go to the halt state.  Once in the halt state, it will stay there indefinitely.
\begin{table}[htbp]\small
  \begin{center}
  \begin{tabular}{cc}
\includegraphics[scale=0.25]{figures/tm4_2-codon_a.png} & \includegraphics[scale=0.25]{figures/tm4_2-codon_b.png}  \\
\includegraphics[scale=0.25]{figures/tm4_2-tRNA_a_l.png} & \includegraphics[scale=0.25]{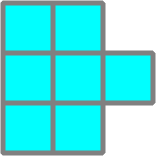}  \\
\\
(L, 0) $\rightarrow$ (L, 0) &	(H, 1) $\rightarrow$ (H, 1) \\
  \end{tabular}
\end{center}
\end{table}

\subsubsection{Infinite Loop}
This is just for curiosity, and again only works for the correct input, but if our input starts with "01" then the following lookup table results in an infinite loop, meaning the tape moves one to the left, then one to the right and so forth.
\begin{table}[htbp]\small
  \begin{center}
  \begin{tabular}{cc}
\includegraphics[scale=0.25]{figures/tm4_2-codon_a.png} & \includegraphics[scale=0.25]{figures/tm4_2-codon_b.png}  \\
\includegraphics[scale=0.25]{figures/tm4_2-tRNA_a_l.png} & \includegraphics[scale=0.25]{figures/tm4_2-tRNA_b_r}  \\
\\
(L, 0) $\rightarrow$ (R, 0) &	(R, 1) $\rightarrow$ (L, 1) \\
  \end{tabular}
\end{center}
\end{table}

\subsection{Turing Machines with no Internal State}
Our simple Turing machine above is not universal, it has only two symbols and basically one state, and it seems to work only for a handful of very special cases.  The symbols are not the problem, but the states are.  However, if we allow for multiple heads on our Turing machine, we can create a Turing machine with an arbitrary number of symbols and states, and it also solves the problems we were seeing above.  Such a Turing machine has no internal state, but instead saves the state on the tape.

We will only sketch here how this could be implemented, using the binary multiply-by-two Turing machine. 
\begin{table}[htbp]\small
  \caption{Times Two}
  \label{tab7_6:tab}
  \begin{center}
  \begin{tabular}{|l|}
\hline
\includegraphics[scale=0.5]{figures/tm_times_two.png}  \\
\hline
Symbols: 0, 1 \\\hline
States: $A_L$, $B_L$ \\\hline
Rules: \\
  ($A_L$, 0) $\rightarrow$ ($A_L$, 0) \\
  ($A_L$, 1) $\rightarrow$ ($B_L$, 0) \\
  ($B_L$, 0) $\rightarrow$ ($A_L$, 1) \\
  ($B_L$, 1) $\rightarrow$ ($B_L$, 1) \\\hline
  \end{tabular}
\end{center}
\end{table}
Here we have two state, A and B, both being left moving.

\subsubsection{States}
We have two heads now, one representing the state, the other the symbol.  As for the state, we use these ant-codons for the input:
\begin{table}[h!]\small
  \begin{center}
  \begin{tabular}{cccccc}
\includegraphics[scale=0.25]{figures/tm4_2-tRNA_a_l.png} & 
\includegraphics[scale=0.25]{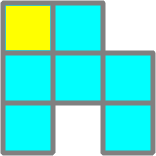} & 
\includegraphics[scale=0.25]{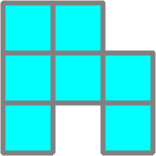} &
\includegraphics[scale=0.25]{figures/tm4_2-tRNA_b_l.png} & 
\includegraphics[scale=0.25]{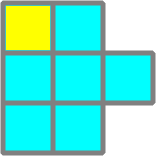} & 
\includegraphics[scale=0.25]{figures/tm4_2-tRNA_b_h.png} \\
\\
State A left & State A right & State A halt & State B left & State B right & State B halt \\
  \end{tabular}
\end{center}
\end{table}

and the following codons for the output:
\begin{table}[h!]\small
  \begin{center}
  \begin{tabular}{cc}
\includegraphics[scale=0.25]{figures/tm4_2-codon_a.png} & 
\includegraphics[scale=0.25]{figures/tm4_2-codon_b.png} \\
\\
State A & State B \\
  \end{tabular}
\end{center}
\end{table}

As is normal for Turing states, once left always left, once right, always right, once halt, always halt.

\subsubsection{Symbols}
As for the symbol, we use these ant-codons for the input:
\begin{table}[h!]\small
  \begin{center}
  \begin{tabular}{cc}
\includegraphics[scale=0.25]{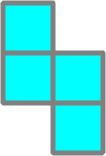} & 
\includegraphics[scale=0.25]{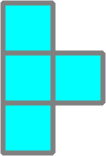} \\
\\
Symbol 0 & Symbol 1 \\
  \end{tabular}
\end{center}
\end{table}

and the following codons for the output:
\begin{table}[h!]\small
  \begin{center}
  \begin{tabular}{cc}
\includegraphics[scale=0.25]{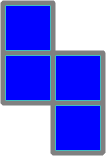} & 
\includegraphics[scale=0.25]{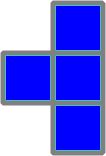} \\
\\
Symbol 0 & Symbol 1 \\
  \end{tabular}
\end{center}
\end{table}

The symbols are independent of the states.  We could choose a 2-state 3-symbol machine where we use 2-codons for the state and 3-codons for the symbol to implement the (2,3) Wolfram-Smith weakly universal Turing machine \cite{smith2007universality}, or one of the (6,4), (5,5), or (4,6) universal Turing machines \cite{rogozhin1996small,neary2009four} with 4-codons for state and symbols.

\subsubsection{Rules}
With the convention the state coming first, the symbol second, we can implement the above rules in the following lookup table:
\begin{table}[htbp]\small
  \begin{center}
  \begin{tabular}{cccc}
\includegraphics[scale=0.25]{figures/tm4_2-codon_a.png} \includegraphics[scale=0.25]{figures/tm4_2-codon_a_short.png} & \includegraphics[scale=0.25]{figures/tm4_2-codon_b.png} \includegraphics[scale=0.25]{figures/tm4_2-codon_a_short.png} & \includegraphics[scale=0.25]{figures/tm4_2-codon_a.png} \includegraphics[scale=0.25]{figures/tm4_2-codon_b_short.png} & \includegraphics[scale=0.25]{figures/tm4_2-codon_b.png} \includegraphics[scale=0.25]{figures/tm4_2-codon_b_short.png} \\

\includegraphics[scale=0.25]{figures/tm4_2-tRNA_a_l.png} \includegraphics[scale=0.25]{figures/tm4_2-tRNA_a_short.png} & \includegraphics[scale=0.25]{figures/tm4_2-tRNA_a_l.png} \includegraphics[scale=0.25]{figures/tm4_2-tRNA_b_short.png} & \includegraphics[scale=0.25]{figures/tm4_2-tRNA_b_l.png} \includegraphics[scale=0.25]{figures/tm4_2-tRNA_a_short.png} & \includegraphics[scale=0.25]{figures/tm4_2-tRNA_b_l.png} \includegraphics[scale=0.25]{figures/tm4_2-tRNA_b_short.png} \\
\\
($A_L$, 0) $\rightarrow$ ($A_L$, 0) &	($A_L$, 1) $\rightarrow$ ($B_L$, 0) &	($B_L$, 0) $\rightarrow$ ($A_L$, 1) &	($B_L$, 1) $\rightarrow$ ($B_L$, 1) \\
  \end{tabular}
\end{center}
\end{table}

Now codons are five blocks long, so the Turing machine has to move by five for each match, and it will always replace the old state/symbol pair with a new state/symbol pair.  Only the above codons are allowed, no others.  And these four 5-codons represent the program of our Turing machine, in this case, multiply by two.

\subsection{Binary Incrementer with 3-Headed Turing Machine}
With the above it is clear how to implement state and symbols in our system.  What we would like to show next is how a state-less Turing machine works.  For this we consider the example of the binary incrementer, implemented with a 3-headed stateless Turing machine.  Recall the binary incrementer:
\begin{table}[htbp]\small
  \caption{Binary incrementer}
  \label{tab7_7a:tab}
  \begin{center}
  \begin{tabular}{|l|}
\hline
\includegraphics[scale=0.5]{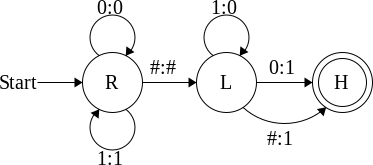}  \\
\hline
Symbols: 0, 1, \# \\\hline
States: $R, L, H$ \\\hline
Rules: \\
  $(R, 0) \rightarrow (R, 0)$  \\
  $(R, 1) \rightarrow (R, 1)$  \\
  $(R, \#) \rightarrow (L, \#)$  \\
  $(L, 0) \rightarrow (H, 1)$  \\
  $(L, 1) \rightarrow (L, 0)$  \\
  $(L, \#) \rightarrow (H, 1)$  \\ \hline
  \end{tabular}
\end{center}
\end{table}
 
To show its workings we consider the binary input "011", with the most significant bit to the left.  We encode it on the tape as "\#011\#" and we initialize the state to be in the R-state.  Our tape represents state-symbol pairs as a tuple "\_0" where the first is the state (could be 'R', 'L', 'H', or underline '\_' denoting undetermined) and the second the symbol (could be '0', '1', '\#').  Thus our initial state "\#011\#" is described by "\_\#R0\_1\_1\_\#".  Since it is a 3-headed machine, it will read the state in the center, but it can write to states in the center, the left, and the right.  We indicate the influence of the head with square brackets.  Hence, initially, the machine sees "[\_\#R0\_1]", which means it is in the R-state, and the current symbol read is the '0'.  For this the rule "$(R, 0) \rightarrow (R, 0)$" is to be applied, meaning stay in the R-state, write out a '0' at the current position, and move to the right.  Since we are moving to the right, our head will shift by one tuple to the right.  Since our new state will be read from the center tuple, we need to make sure it is set to the new state 'R'.  The following depicts this in detail for the "011" example above:
\begin{verbatim}
_#_0_1_1_#_# -> 
[_#R0_1]_1_#_# -> _#[R0R1_1]_#_# -> _#R0[R1R1_#]_# -> _#R0R1[R1R#_#] -> 
_#R0[R1L1R#]_# -> _#[R0L1L0]R#_# -> [_#L0L0]L0R#_# -> [_#H1L0]L0R#_# -> 
_#_1_0_0_#_#
\end{verbatim}

\hfill \break

We can summarize the rules to convert any triplet to another triplet by the following table:
\begin{table}[h!]\small
  \caption{Lookup table 3-headed Turing machine}
  \label{tab7_7:tab}
  \begin{center}
  \begin{tabular}{|l|}
\hline
Rules: \\
  $[$\_\_R0\_\_$]$ $\rightarrow$ $[$\_\_R0R\_$]$ \\
$[$\_\_R1\_\_$]$ $\rightarrow$ $[$\_\_R1R\_$]$ \\
$[$\_\_R\#\_\_$]$ $\rightarrow$ $[$L\_L\#\_\_$]$ \\
$[$\_\_L0\_\_$]$ $\rightarrow$ $[$\_\_H1\_\_$]$ \\
$[$\_\_L1\_\_$]$ $\rightarrow$ $[$L\_L0\_\_$]$ \\
$[$\_\_L\#\_\_$]$ $\rightarrow$ $[$\_\_H1\_\_$]$ \\
  \hline
  \end{tabular}
\end{center}
\end{table}

With this interpretation and lookup-table the operational steps involved are: read the current state and symbol, including its neighbors to the left and right, in the lookup table find what to write to the tape, write it to the tape, and move by one tuple to the left or right.

Our implementation may not be the most efficient one.  Considering that the underline in the above lookup table stands as a wildcard for three states or three symbols this actually is a monster lookup table with 3*3*3*3*6 = 486 entries.  And considering that the matching is a trial-and-error approach, the matching would take forever.  Hence, a stateless Turing machine, although easy to construct, is of no practical value.  However, as we indicated, it is possible to construct efficient Turing machines, even with state.

Our point was to show that the machinery used for self-replication can also be used to construct Turing machines.  Basically, a Turing machine is a copier with two additional components: it can write and it can move in two directions.  
Really surprising is the role of the tRNA: looking at our toy Turing machine, we must realize that its program is the lookup table, which in turn is realized by tRNA.  One could say that the main point of RNA is storage and that of tRNA computation.

\section{Self-Replicating Turing Machine}

We are now ready to look at the bigger picture: we want to suggest how a self-replicating Turing machine could be constructed out of our toy model.  From von Neumann's persepective, it consists of automata A, B, and C, as well as an automaton D, which in this case happens to be a Turing machine.  Let us talk about some of the details.

\subsection{Copying}
The copier takes as input a given mRNA as well as codons/anticodons and produces a copy of the given mRNA.
\begin{figure}[h]
\centering
\includegraphics[scale=0.5]{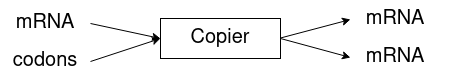}
\caption{Copying of mRNA}
\label{fig8_1:fig}
\end{figure}
An interesting question is where do the codons come from?  From our previous elobarations, one might expect the builder, but that will not work.  To create one 2-codon, for instance, we need three tRNAs, hence at least three 2-codons in the mRNA for description.  Since it is the mRNA we want to copy, it is obvious that this will not work.  However, it is very easy to construct a 2-Codon-Maker, that does nothing but create 2-codons out of normal blocks.
\begin{figure}[h]
\centering
\includegraphics[scale=0.5]{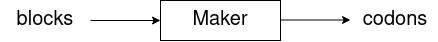}
\caption{Making of Codons}
\label{fig8_2:fig}
\end{figure}

As for the 4-codons, let us point out, that the following 4-codons can be build out of 2-codons, simply by gluing two 2-codons together:
\begin{table}[h!]\small
  \begin{center}
  \begin{tabular}{cccc}
\includegraphics[scale=0.25]{figures/2_4-codons_b.png} & \includegraphics[scale=0.25]{figures/2_4-codons_c.png} & \includegraphics[scale=0.25]{figures/2_4-codons_d.png} & \includegraphics[scale=0.25]{figures/2_4-codons_e.png}   \\
B & C & D & E \\
  \end{tabular}
\end{center}
\end{table}
Hence, restricting ourselves to these 4-codons only, has two advantages: first, they are very easy to build.  Second, the copier, only needs to be a 2-copier, and we never really need any 4-codons during the copy process, hence there also is no need for gluing them together.  A disadvantage might be the fact that mutations are now more difficult.  But since we are building a self-replicating and not a self-reproducing Turing machine \cite{merkle2004kinematic}, this is no issue.

\subsection{Construction}
\begin{figure}[h]
\centering
\includegraphics[scale=0.5]{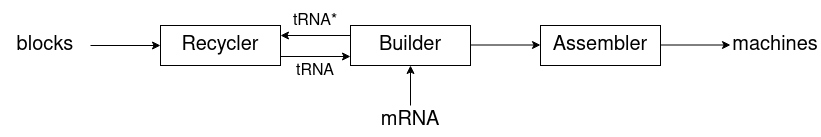}
\caption{Construcing Machines}
\label{fig8_3:fig}
\end{figure}

The constructor, consisting of builder and assembler, takes mRNA as instruction, and with the help of tRNA constructs machines (Figure \ref{fig8_3:fig}).  In the process, mRNA is not modified at all.  As for the tRNA we have already indicated previously, that we need to recycle spent tRNA.  Creating tRNA directly through the builder, makes as little sense as creating codons through the builder.  However, since tRNA are recycled, we do not need to have a special tRNA-Maker, instead we could build one or two copies of each tRNA, and use them in the recycling process. 

\subsection{Self-Replication}
\begin{figure}[h]
\centering
\includegraphics[scale=0.5]{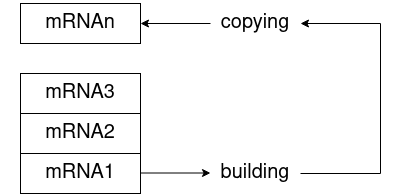}
\caption{Self-replication}
\label{fig8_4:fig}
\end{figure}

One way the self-replication could happen is through the self-replication cycle indicated in figure \ref{fig8_4:fig}.  We have a stack of different mRNAs. 
For instance, mRNA1 might encode the tRNAs, mRNA2, might be the instructions to build the builder, etc.  Each mRNA is first sent to the constructor, after which it is forwarded to the copier, which makes a copy, and then is returned back to the stack of mRNAs.  The different mRNAs must contain at a minimum the instructions to build the machines and components needed for the replication process, but there are additional mRNAs describing how to construct a Turning machine. 

In parallel to the replication cycle, an existing Turing machine might be doing its calculations.  An interesting thought is where the program, that is the tRNA of the child Turing machine, comes from.  Most likely it would also be created by the self-replication cycle, if one wanted the child Turing machine to run the same program as the parent.  However, one might actually want to change the programming of the child Turing machine, then the tRNA of the child Turing machine would need to be created by some other process, maybe even introducing mutations.  More interesting might be the tape: assume the parent Turing machine has done some lengthy calculation and arrived at an interesting result, which it might actually want to pass on to the child Turing machine.  This could be accomplished, if part or all of the parents tape is copied or transferred to the childs tape.  This would most certainly allow for some very interesting scenarios.

\section{Conclusion}

Let us recall our assumptions: basic building blocks that can be glued together to form larger compounds, with no specific assumptions about their material properties.  And a mover block that can expand to twice its size with a simple form of time ordering.  Critical is also three-dimensionality, although, as von Neumann noted, a multiply-connected plane would also do.
Naturally, energy is important, and the fact that the mover can move a relatively large amount compared to its own size.

\subsection{Bits of Information}
These simple rules lead to a "thermodynamically highly improbable scenario" as von Neumann called it \cite{neumann1966theory}.  However, because of the simplicity of these rules, we actually tend to think that, maybe, it is not so improbable after all.  
Our simple rules lead inevitably to self-reproduction, elucidating why for roughly half of the age of the universe, life has been thriving on this planet, and most likely others.

H. Jacobson in his famous paper "Information, reproduction, and the origin of life" \cite{jacobson1955information} made various assumptions and predictions about evolution.  One such prediction was that the simplest living beings could have at most about 200 bits of information for evolution to ever occur in the time frame and conditions found on Earth.  For our simple machines, we can actually calculate their information content.  We take our machine description language (details are in the appendix) and compress it.  We simply replace the "\_\_\_", "b\_\_", etc., by numbers.  Since 0 to 7 is enough, 3-bits for the encoding is all that is needed.  The copier has 45 characters, that is 135 bits of information, and the builder has 71 characters, that is 213 bits of information.  Hence, both within reach of evolution.

\subsection{Von Neumann's five Questions}

Von Neumann was interested in "organizational questions about complicated organisms" \cite{neumann1966theory}.  
We can revisit some of his questions, and try to answer them from our toy theory point of view.

\textbf{A) Logical universality:} when is a class of automata logically universal, and is a single automaton logically universal?  Since we were able to construct a Turing machine, the answer to the first is affirmative.  For the second, it depends on what one calls an automaton, since most of our machines are not logically universal.

\textbf{B) Constructability:} can an automaton be constructed by another automaton, and what class of automata can be constructed by a single suitable automaton?  Our builder can construct all necessary parts for all our machines.  However, we have only indicated how the parts are assembled to a whole, and left out a few steps.  So we believe we can, but we have not shown this.

\textbf{C) Construction universality:} is any single automaton construction universal?  According to our definition of automaton, the answer is no.

\textbf{D) Self-reproduction:} is there a self-reproducing automaton, and is there an automaton which can both reproduce itself and perform further tasks?  Also here the answer is affirmative: our builder can produce any parts needed for reproduction, but in addition can produce arbitrary other parts, including parts for a Turing machine.

\textbf{E) Evolution:} can the construction of automata by automata progress from simpler types to increasingly complicated types, and assuming some suitable definition of efficiency, can this evolution go from less efficient to more efficient automata?  Here we can not really add much more than von Neumann did (sub chapter 1.8 \cite{neumann1966theory}).  If we consider our toy mechanism in the context of biological evolution, we are basically at the stage of the primordial soup.  There are no cell walls, no individuum exist, the advent of species has not happened yet.

\subsection{Missing Pieces}
We have shown a major part of the universal constructor A and the universal copier B.  Automaton D is trivial, because that could be any machine, only the possibility of mutation is required. 

What we have not touched upon is automaton C.  It seems to be significantly more complex, than what von Neumann seems to have indicated, at least for our mechanism.  Among many things it has to transport mRNA to the proper location, start and stop the copy process, move and assemble components that the builder and assembler produce, initiate and control shredding and recycling, increase or decrease density of required components, identify and sort components, build highways and walls, adjust the rate of mutation, and many more things.

Also, we have not shown where energy comes from, how it is stored, or a mechanism for the internal timing.

\subsection{Final Remarks}
Interesting is with which ease we arrive at a physical implementation of a simple Turing machine.  Although considering that Turing machines can actually be found in nature \cite{benenson2012biomolecular,varghese2015molecular} this may not be so surprising after all.  Clearly within reach is a self-replicating Turing machine, which gives the word "physical computation" a somewhat new meaning. 
Although this is only a toy mechanism, it helps us to learn about the interconnection between life, computation and language.

\section{Acknowledgement}
I would like to thank W.Z. Taylor for discussions, feedback and suggestions.  In addition, I am grateful to an anonymous referee for very helpful suggestions.

\section{Appendix}

The animation sequences presented in this paper are also available as short movie sequences.  They were created with the physical simulation program written in Java, included with this paper.  

All machines are described by our machine description language (MDL).  It is a simple description of the three dimensional positions of its components.  We denote normal blocks by 'b', mover blocks by 'M', and gluer blocks by 'G'.  In addition, for mover and gluer blocks we need to indicate the direction in which they are pointing, which is a number between 0 and 5, where 0 indicates positive x-direction, 1 positive y-direction and so forth.  And for the mover we also need to indicate timing, a number between 0 and 9, indicating the relative order of timing, meaning, 0 is before 1, and so forth.  Hence, three characters describe one position.  If there is no block at a given position, this is indicated by three underlines.  For instance, "M04" indicates a mover, pointing in positive x-direction, which will expand at time tick four.

As an example, consider the MDL for the builder.
\begin{lstlisting}[frame=single] 
// z=0:
b_____M25
b_____M25
______M25

// z=1:
b________
_________

// z=2:
M04______
_________

// z=3:
b________
_________

// z=4:
b__G01___
___G01___
\end{lstlisting}
One line describes the relative position of the blocks in x-direction, a new line in the y-direction, and an empty line indicates the start of a new plane in z-direction.  This completely describes our machines.

\begin{figure}[h]
\centering
\subfloat[3D tRNA]{
\includegraphics[scale=0.25]{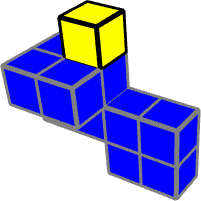}
\label{fig5_28a:sf1}}
\qquad
\subfloat[Glue component]{
\includegraphics[scale=0.25]{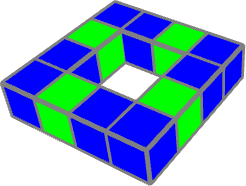}
\label{fig5_28a:sf2}}
\caption{3D Builder}
\label{fig5_28a:fig}
\end{figure}

Finally, a remark on the restriction to printing only 2-by-n structures.  It is possible to print 3D components.  For instance, adding a little "3D platform" to the tRNA and modifying the glue component of the builder as indicated in figure \ref{fig5_28a:fig} would be one possibility.

\nocite{*}
\bibliographystyle{fundam}
\bibliography{citations}

\begin{thebibliography}{10}
\providecommand{\url}[1]{\texttt{#1}}
\providecommand{\urlprefix}{URL }
\expandafter\ifx\csname urlstyle\endcsname\relax
  \providecommand{\doi}[1]{doi:\discretionary{}{}{}#1}\else
  \providecommand{\doi}{doi:\discretionary{}{}{}\begingroup
  \urlstyle{rm}\Url}\fi
\providecommand{\eprint}[2][]{\url{#2}}

\bibitem{neumann1966theory}
Neumann Jv.
\newblock Theory of self-reproducing automata.
\newblock \emph{Edited by Arthur W. Burks}, 1966.

\bibitem{rocha2015neumann}
Rocha L.
\newblock Von Neumann and Natural Selection.
\newblock \emph{Biologically Inspired Computing Lecture Notes}, 2015.

\bibitem{o2013principles}
O’Donnell M, Langston L, Stillman B.
\newblock Principles and concepts of DNA replication in bacteria, archaea, and
  eukarya.
\newblock \emph{Cold Spring Harbor perspectives in biology}, 2013.
\newblock \textbf{5}(7):a010108.

\bibitem{alberts2017molecular}
Alberts B.
\newblock Molecular biology of the cell.
\newblock WW Norton \& Company, 2017.

\bibitem{merkle2004kinematic}
Freitas~Jr RA, Merkle RC.
\newblock Kinematic Self-Replicating Machines.
\newblock Landes BioScience, 2004.

\bibitem{turing1936computable}
Turing AM, et~al.
\newblock On computable numbers, with an application to the
  Entscheidungsproblem.
\newblock \emph{J. of Math}, 1936.
\newblock \textbf{58}(345-363):5.

\bibitem{minsky1967computation}
Minsky M.
\newblock Computation: Finite and Infinite Machines Prentice-Hall, Inc.
\newblock \emph{Englewood Cliffs, N. J.(also London 1972)}, 1967.

\bibitem{smith2007universality}
Smith A.
\newblock Universality of Wolfram’s 2, 3 Turing machine.
\newblock \emph{Complex Systems}, 2007.

\bibitem{rogozhin1996small}
Rogozhin Y.
\newblock Small universal Turing machines.
\newblock \emph{Theoretical Computer Science}, 1996.
\newblock \textbf{168}(2):215--240.

\bibitem{neary2009four}
Neary T, Woods D.
\newblock Four small universal Turing machines.
\newblock \emph{Fundamenta Informaticae}, 2009.
\newblock \textbf{91}(1):123--144.

\bibitem{jacobson1955information}
Jacobson H.
\newblock Information, reproduction and the origin of life.
\newblock \emph{American Scientist}, 1955.
\newblock \textbf{43}(1):119--127.

\bibitem{benenson2012biomolecular}
Benenson Y.
\newblock Biomolecular computing systems: principles, progress and potential.
\newblock \emph{Nature Reviews Genetics}, 2012.
\newblock \textbf{13}(7):455--468.

\bibitem{varghese2015molecular}
Varghese S, Elemans JA, Rowan AE, Nolte RJ.
\newblock Molecular computing: paths to chemical Turing machines.
\newblock \emph{Chemical science}, 2015.
\newblock \textbf{6}(11):6050--6058.

\end{thebibliography}


\end{document}